\documentclass[useAMS,usenatbib,twocolumn]{mnras}
\usepackage{graphicx}
\usepackage{amsmath}
\usepackage{amssymb}
\usepackage{xspace}
\usepackage{commath}
\usepackage{times}
\usepackage{bm} 
\usepackage{balance}
\usepackage{hyperref}
\usepackage{mathtools}
\usepackage{stmaryrd}

\hypersetup{dvips, colorlinks=true, linkcolor=black, citecolor=black, filecolor=blue, urlcolor=black}

\usepackage{acronym}
\newacro{PDF}{probability distribution function}
\newacroplural{PDFs}{probability distribution functions}
\newcommand{\PDF}{\ac{PDF}}
\newcommand{\PDFs}{\acp{PDF}}
\newacro{DF}{distribution function}
\newacroplural{DFs}{distribution functions}
\newcommand{\DF}{\ac{DF}}
\newcommand{\DFs}{\acp{DF}}
\newacro{BH}{black hole}
\newacroplural{BHs}{black holes}
\newcommand{\BH}{\ac{BH}}
\newcommand{\BHs}{\acp{BH}}
\newacro{IMBH}{intermediary mass black hole}
\newacroplural{IMBHs}{intermediary mass black holes}
\newcommand{\IMBH}{\ac{IMBH}}
\newcommand{\IMBHs}{\acp{IMBH}}
\newacro{FP}{Fokker-Planck}
\newcommand{\FP}{\ac{FP}}
\newacro{SRR}{scalar resonant relaxation}
\newcommand{\SRR}{\ac{SRR}}
\newacro{VRR}{vector resonant relaxation}
\newcommand{\VRR}{\ac{VRR}}
\newacro{NR}{non-resonant relaxation}
\newcommand{\NR}{\ac{NR}}
\newacro{RR}{resonant relaxation}
\newcommand{\RR}{\ac{RR}}
\newacro{LR}{likelihood ratio}
\newcommand{\LR}{\ac{LR}}

\usepackage{soul} \usepackage{xcolor}
\definecolor{aquamarine}{rgb}{0.5, 1.0, 0.83}

\newcommand{\rd}{\mathrm{d}}

\newcommand{\Pth}{P_{\mathrm{th}}}
\newcommand{\DRR}{D^{\mathrm{RR}}}
\newcommand{\DNR}{D^{\mathrm{NR}}}

\newcommand{\np}{n^{\prime}}
\newcommand{\ap}{a^{\prime}}
\newcommand{\jp}{j^{\prime}}
\newcommand{\Mp}{M^{\prime}}
\newcommand{\Min}{\mathrm{Min}}
\newcommand{\Max}{\mathrm{Max}}
\newcommand{\rp}{r^{\prime}}
\newcommand{\mO}{\mathcal{O}}
\renewcommand{\fp}{f^{\prime}}
\newcommand{\Jc}{J_{\mathrm{c}}}
\newcommand{\Ftot}{F_{\mathrm{tot}}}
\newcommand{\nup}{\nu_{\mathrm{p}}}
\newcommand{\nuGR}{\nu_{\mathrm{GR}}}
\newcommand{\nuM}{\nu_{\mathrm{M}}}
\newcommand{\nuKep}{\nu_{\mathrm{Kep}}}
\newcommand{\onuM}{\overline{\nu}_{\mathrm{M}}}
\newcommand{\hM}{h_{\mathrm{M}}}
\newcommand{\Card}{\mathrm{Card}}
\newcommand{\gp}{g^{\prime}}
\newcommand{\rg}{r_{\mathrm{g}}}
\newcommand{\MBH}{M_{\bullet}}
\newcommand{\jlc}{j_{\mathrm{lc}}}
\newcommand{\apmin}{a^{\prime}_{\mathrm{min}}}
\newcommand{\apmax}{a^{\prime}_{\mathrm{max}}}
\newcommand{\Kres}{K_{\mathrm{res}}}
\newcommand{\rh}{r_{\mathrm{h}}}
\newcommand{\hbr}{\widehat{\mathbf{r}}}
\newcommand{\Jz}{J_{z}}
\newcommand{\hbJ}{\widehat{\mathbf{J}}}
\newcommand{\br}{\mathbf{r}}
\newcommand{\bv}{\mathbf{v}}
\renewcommand{\balpha}{\bm{\alpha}}
\newcommand{\erf}{\mathrm{erf}}
\newcommand{\mL}{\mathcal{L}}
\newcommand{\mpc}{\rm{mpc}}
\newcommand{\ellmax}{\ell_{\mathrm{max}}}
\newcommand{\mLmax}{\mathcal{L}_{\mathrm{max}}}
\newcommand{\nobs}{n_{\mathrm{obs}}}
\newcommand{\yr}{\mathrm{yr}}
\newcommand{\kyr}{\mathrm{kyr}}
\newcommand{\Myr}{\mathrm{Myr}}
\newcommand{\Gyr}{\mathrm{Gyr}}

\begin{document}

\title[Resonant Relaxation of S-stars]{Mapping the Galactic centre's dark cluster
via Resonant Relaxation}

\author[K. Tep, et al.]{Kerwann Tep$^{1}$, Jean-Baptiste Fouvry$^{1}$, Christophe Pichon$^{1,2,3}$, 
Gernot Hei\ss el$^4$, \newauthor Thibaut Paumard$^4$, Guy Perrin$^4$ and 
Frederic Vincent$^4$ \\
\noindent
$^{1}$ CNRS and Sorbonne Universit\'e, UMR 7095, Institut d'Astrophysique de Paris, 98 bis Boulevard Arago, F-75014 Paris, France\\
$^{2}$ Korea Institute for Advanced Study, 85 Hoegiro, Dongdaemun-gu, 02455 Seoul, Republic of Korea\\
$^{3}$ IPhT, DRF-INP, UMR 3680, CEA, Orme des Merisiers Bat 774, 91191 Gif-sur-Yvette, France\\
$^4$ LESIA, Observatoire de Meudon, 5 Place Jules Janssen, 92190 Meudon, France
}

\maketitle

\begin{abstract}
Supermassive black holes in the centre of galaxies
dominate the gravitational potential of their surrounding stellar clusters.
In these dense environments, stars follow nearly Keplerian orbits,
which get slowly distorted as a result of the potential fluctuations
generated by the stellar cluster itself. 
In particular, stars undergo a rapid relaxation of their eccentricities
through both resonant and non-resonant processes.
An efficient implementation of the resonant diffusion coefficients
allows for detailed and systematic explorations
of the parameter space describing the 
 properties of the stellar cluster.
In conjunction with recent observations
of the S-cluster orbiting SgrA*,
this framework can be used
to jointly constrain the distribution of the unresolved, old,
background stellar cluster and the characteristics of a putative dark cluster.
Specifically, we show how this can be used to estimate the typical mass and cuspide exponent of intermediate-mass black holes consistent with the relaxed state
of the distribution of eccentricities in the observed S-cluster.
This should prove useful in constraining super massive black hole formation scenarios.
\end{abstract}


\section{Introduction}
\label{sec:intro}

Galactic nuclei are among the densest stellar systems
of the universe.
 Recent outstanding observations
keep providing us with new information on these regions.
These include detailed census of stellar populations
around SgrA* in the centre of the Galaxy~\citep{Ghez2008,Gillessen2017},
the first observation of the relativistic precession
of the star S2 within our own Galactic centre~\citep{Gravity2020};
the observation of a cool accretion disc around SgrA*~\citep{Murchikova2019};
the first image of the shadow of M87~\citep{EHT2019}
and mergers of binary black holes recently detected via gravitational wave emission~\citep{Abbott+2019},
that may or may not have occurred in galactic nuclei. 
These various successes will soon be complemented
with ever finer resolution around SgrA*~\citep{Gravity2017},
as well as much larger stellar populations
permitted by the planned upgrade on
VLTI/GRAVITY~\citep{Gravity+2019,Gravity2021},
as well as the upcoming thirty-meter class telescopes
such as TMT~\citep{Do2019}
and ELT/MICADO~\citep{davies2018micado,pott2018micado}.
Such a wealth of observational information
offers new venues to investigate the details
of stellar dynamics around supermassive \BHs\@,
as well as probe the possible presence of \IMBHs\
in these regions~\citep{PortegiesZwart2002}.

Indeed, in the crowded region of galactic nuclei,
the gravitational potential remains nonetheless 
dominated by the central supermassive \BH\@.
Because of the steep gravitational potential
that this \BH\ induces, galactic nuclei involve a wide range
of relaxation processes and timescales~\citep{RauchTremaine1996,HopmanAlexander2006,Merritt2013,Alexander2017}.
These are:
(i) the dynamical time associated with the fast Keplerian motion
imposed by the central \BH\@
(${ \sim 16 \, \yr }$ for S2, e.g.\@,~\cite{Gillessen2017}).
On longer timescales, one can formally smear out the stars
along their stellar orbits so that they are effectively replaced
with massive eccentric wires;
(ii) the in-plane precession time of the Keplerian
wires generated by both the relativistic corrections
from the central \BH\ and the stellar mean potential
(${ \sim 3\!\times\! 10^{4} \, \yr }$ for S2, e.g.\@,~\cite{Gravity2020});
(iii) the \VRR\
time~\citep[see, e.g.\@,][]{KocsisTremaine2015,Fouvry+2019}
during which, as a result of the non-spherically symmetric stellar fluctuations
and relativistic corrections induced by a spinning \BH\@,
stars undergo a stochastic reshuffling of their orbital orientations
(${ \sim 1 \, \Myr }$ for S2, e.g.\@,~\cite{KocsisTremaine2011});
(iv) the \SRR\ time~\citep{RauchTremaine1996,BarOrAlexander2016,SridharTouma2016,BarOrFouvry2018}
during which resonant torques between the in-plane precessing wires
lead to a diffusion of the wires' eccentricities
(${ \sim 10 \, \Myr }$ for S2, e.g.\@,~\cite{BarOrFouvry2018});
(v) the \NR\ time~\citep{BahcallWolf1976,LightmanShapiro1977,BarOrAlexander2013,Vasiliev2017} during which nearby pairwise scatterings
slowly drive the long-term relaxation of the wires' semi-major axes,
as well as of their eccentricities
(${ \sim 1 \, \Gyr }$ for S2, e.g.\@,~\cite{KocsisTremaine2011}).

The leverage provided by modelling these dynamical processes with
 recent observations should allow us to constrain hidden features of the galactic centre.
Here, we will focus
on the relaxation of stellar eccentricities
in galactic nuclei, through processes (iv) and (v).
Indeed, as emphasised in~\cite{Gillessen2017}
(see Fig.~{13} therein),
the S-cluster orbiting SgrA* (for ${ a \simeq 5 \, \mpc }$)
has been observed with a significantly relaxed distribution
of eccentricities.
Detailed spectroscopic observations~\citep{Habibi2017} also
provide us with well constrained main-sequence
ages for these same stars.
The age of the young B-stars at a distance of ${ \sim 10 \, \mpc }$
from the \BH\ is comparable to the resonant relaxation time,
implying that resonant relaxation may have played an important role in their dynamical structure~\citep{HopmanAlexander2006}.
Hence,
any credible diffusion mechanism
has to be efficient enough to drive a significant eccentricity relaxation
of the S-stars within their lifetime.
Equivalently, one may use this constraint
as a dynamical probe to further characterise
the properties of the unresolved old stellar and putative dark cluster,
which both drive the relaxation of the S-stars themselves \citep[see, e.g.][]{Generozov2020}.
Following \cite{Merritt2009, Antonini2012},
Monte Carlo simulations have similarly shown that stellar mass \BHs\
can reduce the resonant relaxation time near the present-day location of the S-stars to ${ \sim 10 \, \Myr }$, which is of the order of the age
of the S-stars~\citep{Habibi2017}.
This is the purpose of this paper.
We use kinetic theory to constrain
the range of cluster models for SgrA*
that are compatible with the observational requirement
of having significantly relaxed S-stars eccentricities.

The paper is organised as follows.
In section~\ref{sec:Relaxation},
we briefly review the two main dynamical processes
in galactic nuclei
through which stars can relax in eccentricities.
We also detail our (fast) numerical computations
of the associated diffusion coefficients.
In section~\ref{sec:Application},
we place first constraints on the stellar distribution
of the unresolved old (stellar and dark) cluster
using these diffusion process in conjunction
with recent observations of the S-cluster. 
We also present a fiducial model 
in which we can vary the number of stars to anticipate what upcoming instruments will be able to measure.
Finally, we discuss these results
and conclude in section~\ref{sec:conclusion}.

\section{Long-term relaxation}
\label{sec:Relaxation}

\subsection{Mean-field dynamics}
\label{sec:MeanField}

Let us consider a test star
orbiting within a galactic nuclei
containing a central supermassive \BH\
of mass $\MBH$.
Because the potential is dominated by the central \BH\@,
this test star follows a (nearly) Keplerian orbit
that we can describe using orbital elements~\citep{MurrayDermott1999}
written as ${ (M , \omega , \Omega , \Jc , J , \Jz) }$.
In these notations, the dynamical angles are $M$ the mean anomaly,
i.e.\ the location of the star along its Keplerian orbit,
$\omega$ the argument of the pericentre,
and $\Omega$ the longitude of the ascending node.
The associated actions are given by
\begin{equation}
\Jc = \sqrt{G \MBH a} ;
\quad
J = \Jc \sqrt{1 - e^{2}} ;
\quad
\Jz = J \cos (I) .
\label{def_actions}
\end{equation}
In that expression, $\Jc$ is the circular angular momentum,
$a$ the orbit's semi-major axis and $e$ its eccentricity,
$J$ the magnitude of the angular momentum vector,
$I$ its inclination,
and $\Jz$ its projection along the $z$-axis.

The fast Keplerian motion of the star is
then described by ${ \dot{M} \!=\! \nuKep }$,
with the Keplerian frequency
\begin{equation}
\nuKep (a) = \sqrt{\frac{G \MBH}{a^{3}}} .
\label{def_nuKep}
\end{equation}
This dynamical time being so short,
the traditional approach of secular dynamics
is to smear out the stars along their Keplerian
ellipses~\citep[see, e.g.\@,][]{ToumaTremaine2009},
so that they formally become massive wires.
Following this orbit-average over $M$,
the conjugate coordinate $\Jc$ (and therefore $a$)
is conserved by adiabatic invariance
for the secular dynamics.

Describing the long-term dynamics
of stellar orbits in galactic nuclei
amounts then to describing the long-term evolution
of the remaining five coordinates ${ \{ \omega , \Omega , \Jc , J , \Jz \} }$.
On longer timescales,
the wires undergo some in-plane precession,
described by
\begin{equation}
\frac{\rd \omega}{\rd t} = \nup (a , j) = \nuGR (a , j) + \nuM (a , j) ,
\label{def_nup}
\end{equation}
Here, ${ \nup (a,j) }$ describes the total precession frequency
of the wire's pericentre.
It is given by the joint contribution
from the relativistic corrections from the central
\BH\@,
i.e.\ the Schwarzschild precession~\citep{Merritt2013},
through the term ${ \nuGR (a, j) }$,
as well as from the mass precession
imposed by the mean background stellar cluster, ${ \nuM (a,j) }$.
Appendix~\ref{sec:ResCond}
presents explicit expressions of both of these frequencies.

As mentioned in introduction,
on timescales longer than the precession time
the Keplerian wires will be subject
to three main relaxation processes,
namely the \VRR\@, during which
the direction of the stellar orbital plane, ${ \hbJ \!=\! (\Omega , I) }$,
diffuses;
the \SRR\@, during which $J$, i.e. the eccentricity $e$,
diffuses;
and finally, the \NR\@, during which both ${ (\Jc , J) }$
diffuse, i.e.\ wires undergo changes in both $a$ and $e$. 
Here, we are interested in the process of eccentricity relaxation.
Stellar eccentricities
can relax both through \SRR\ and \NR\@,
and we now briefly recall the key properties
of these two processes.

\subsection{Eccentricity relaxation}
\label{sec:EccentricityRelaxation}

We are interested in the dynamics of the S-stars
on timescales of the order ${ \sim 10 \, \Myr }$,
i.e.\ their stellar age.
Since this age is generically much shorter than the timescale for \NR\@,
we may assume that the semi-major axis of each star, $a$,
is conserved.
As a consequence, we keep track of the stars' eccentricities
through the dimensionless angular momentum
\begin{equation}
j = \sqrt{1 - e^{2}} .
\label{def_j}
\end{equation}
Characterising the relaxation of the S-stars
amounts then to describing the long-term dynamics
of their $j$.

This diffusion is sourced by the potential
fluctuations generated by the background unresolved cluster.
This cluster is expected to be old,
i.e.\ has been orbiting around SgrA*
for a time much longer than the \SRR\ relaxation time.
As such, we may assume that it has already 
 fully relaxed all its orbital elements.
We therefore assume that it
has a spherically symmetric distribution of orientations,
and, importantly, follows a thermal distribution of eccentricities.

In that limit, the eccentricities of the test particles,
i.e.\ the eccentricities of the S-stars,
follow a diffusion equation of
the form~\citep{BarOrAlexander2016}
\begin{equation}
\frac{\partial P (j , t \,|\, a)}{\partial t} = \frac{1}{2} \frac{\partial }{\partial j} \bigg[ j \, D_{jj} (a , j) \, \frac{\partial }{\partial j} \bigg( \frac{P (j , t \,|\, a)}{j} \bigg) \bigg] ,
\label{eq:FP_nice}
\end{equation}
where ${ P (j,t \,|\, a) }$ describes the \PDF\
of test stars' eccentricities, $j$,
for a given semi-major axis $a$,
as a function of time,
normalised so that ${ \!\int\! \rd j \, P(j , t \,|\, a) \!=\! 1 }$. The flux is conserved because the boundary conditions are such that there is, by design, no flux that escapes the ${ [0,1] }$ interval. Indeed, at ${ j \!=\! 0 }$, the flux vanishes because of the $j$-factor in equation~\eqref{eq:FP_nice}, while at ${ j \!=\! 1 }$, the diffusion coefficient $D_{jj}$ goes to 0. We neglect the supermassive \BH\@'s loss-cone region, which would drive a small exiting flux at low $j$.
Finally, we also neglect the diffusion in $a$
(hence ${ D_{aa}\!=\!0 }$), which is minor compared to that in angular momentum $j$ \citep{BarOrAlexander2016}.
Stars diffuse therefore at fixed semi-major axes.

The steady state of equation~\eqref{eq:FP_nice} is given
by the thermal solution, ${ \Pth (j \,|\, a) = 2j }$,
i.e.\ the eccentricity \PDF\ also followed
by the background stars.
Let us already note that equation~\eqref{eq:FP_nice}
can be rewritten under the more classical \FP\ form as
\begin{align}
\frac{\partial P (j , t \,|\, a)}{\partial t} = - \frac{\partial }{\partial j} & \, \bigg[ D_{j} (a , j) \, P (j , t \,|\, a) \bigg] 
\nonumber
\\
+ \frac{1}{2} \frac{\partial^{2} }{\partial j^{2}} & \, \bigg[ D_{jj} (a , j) \, P (j , t \,|\, a) \bigg] ,
\label{eq:FP_classical}
\end{align}
where the first- and second-order diffusion coefficients
satisfy the fluctuation-dissipation relation~\citep{BarOrAlexander2016}
\begin{equation}
D_{j} = \frac{1}{2 j} \, \frac{\partial }{\partial j} \bigg[ j \, D_{jj} \bigg] .
\label{fluctuation_dissipation}
\end{equation}
The rewriting from equation~\eqref{eq:FP_classical} is useful
to perform Monte-Carlo integrations of the stochastic dynamics,
as presented in Appendix~\ref{sec:MonteCarlo}.

In equation~\eqref{eq:FP_nice}, we introduced the diffusion coefficient
in angular momentum, ${ D_{jj} (a, j) }$,
that are the sum of two contributions
\begin{equation}
D_{jj} (a , j) = \DRR_{jj} (a , j) + \DNR_{jj} (a , j) ,
\label{sum_Djj}
\end{equation}
where ${ \DRR_{jj} (a,j) }$ captures the contribution from \RR\@,
while ${ \DNR_{jj} (a,j) }$ is associated with the contribution
from \NR\@.
We now detail the content of each of these coefficients.

\subsection{Scalar Resonant Relaxation}
\label{sec:SRR}

A first source of eccentricity relaxation
stems from the long-range resonant couplings between
the in-plane precessing wires.
Following~\cite{BarOrFouvry2018} and references therein,
the \SRR\ diffusion coefficients read
\begin{align}
\hskip -0.2cm
\DRR_{jj} (a , j) \!= \! \frac{4 \pi G^{2}}{\Jc^{2}}\! & \sum_{n = 1}^{+ \infty} \! \sum_{\substack{\np = - \infty \\ \np \neq 0}}^{+ \infty} \!\! \!\!\frac{n^{2}}{|\np|} \times\notag
\\
&\!\! \int \!\! \rd \ap \, \Ftot (\ap\! , \jp) \frac{|A_{n\np} (a , j , \ap \!, \jp)|^{2}}{|\partial_{j} \nup (\ap\! , \jp)|} ,
\label{Djj_SRR}
\end{align}
where ${ \Jc \!=\! \Jc (a) }$ was defined in equation~\eqref{def_actions},
and $\jp$ is the implicit solution of the resonance constraint
\begin{equation}
\nup (\ap , \jp) = \frac{n}{\np} \, \nup (a , j) ,
\label{resonance_constraint}
\end{equation}
with the in-plane precession frequencies, ${ \nup (a,j) }$,
already introduced in equation~\eqref{def_nup}.

In equation~\eqref{Djj_SRR},
we introduced the \DF\@, ${ \Ftot (\ap , \jp) }$,
to describe the background cluster,
whose potential fluctuations are responsible
for the long-term diffusion of stellar eccentricities.
It is defined as
\begin{equation}
\Ftot (a , j) = \sum_{i} m_{i}^{2} N_{i} (a) \, f_{i} (j \,|\, a) ,
\label{def_Ftot}
\end{equation}
where the sum over $i$ runs over all the sub-populations
of the background cluster.
Each population is characterised by an individual mass, $m_{i}$,
while ${ N_{i} (a) }$ is the number of stars per unit semi-major axis $a$,
and ${ f_{i} (j \,|\, a) }$ is the conditional \PDF\ of $j$ for a given $a$,
normalised so that ${ \!\int\! \rd j f_{i} (j \,| \, a) \!=\! 1 }$.
In practice, in order to ease the numerical resolution
of the resonance condition (see Appendix~\ref{sec:ResCond})
and the computation of the \NR\ diffusion coefficients,
we assume that each population follows
a power law distribution in semi-major axes
and is also fully relaxed in eccentricity, i.e.\ ${ f_{i} (j \,|\, a) = 2 j }$,
owing to their old dynamical age.
We further detail all our normalisation conventions in Appendix~\ref{sec:Cusp}.

The resonant diffusion coefficients from equation~\eqref{Djj_SRR}
involve the coupling coefficients
${ |A_{n\np}|^{2} }$
that describe the efficiency of the resonant
coupling between two wires.
They read
\begin{equation}
\big| A_{n \np} (a , j , \ap\! , \jp) \big|^{2} \!=\! 16 \pi^{2} \sum_{\ell} \!\frac{|y_{\ell}^{n}|^{2} |y_{\ell}^{\np}|^{2}}{(2 \ell + 1)^{3}} \, \!\big| K_{n \np}^{\ell} (a , j , \ap\! , \jp) \big|^{2} ,
\label{def_A} \notag
\end{equation}
with the constant coefficients
${ y_{\ell}^{n} \!=\! Y_{\ell}^{n} (\tfrac{\pi}{2} , \tfrac{\pi}{2}) }$,
where the spherical harmonics are defined
with the convention ${ \!\int\! \rd \hbr \, |Y_{\ell}^{n} (\hbr) |^{2} \!=\! 1 }$.
This equation involves the pairwise in-plane coupling coefficients
${ K_{n\np}^{\ell} }$ that read
\begin{equation}
K_{n\np}^{\ell} (a , j , \ap , \jp) \!=\! \bigg\langle\! \cos (n f) \cos (\np \fp) \frac{\Min [r , \rp]^{\ell}}{\Max [r , \rp]^{\ell + 1}} \!\bigg\rangle_{\!\!\circlearrowright} ,
\label{def_Klnnp}
\end{equation}
where $f$ is the true anomaly,
while ${ \langle \,\cdot\, \rangle_{\circlearrowright} }$
stands for the orbit-average over both radial oscillations.
Let us already emphasise that the coupling coefficients
from equation~\eqref{def_A} satisfy various symmetry properties.
First, as imposed by ${ |y_{\ell}^{n}|^{2} }$ and ${ |y_{\ell}^{\np}|^{2} }$,
these coefficients are non-zero only when ${ |n| , |\np| \leq \ell }$,
as well as ${ (\ell \!-\! n) }$ and ${ (\ell \!-\! \np) }$ even.
In addition, we note that we have
${ |A_{n\np}|^{2} \!=\! |A_{\pm n \pm \np}|^{2} }$,
i.e.\ the strength of the ${ (n , \np) }$ coupling
is independent of the sign of the resonance numbers.
These are all important features which will allow us to reduce the required number
of evaluations of the coupling coefficients.
Finally, in practice, in Eq.~\eqref{def_A},
we truncate the harmonics up to a given $\ellmax$.

In equation~\eqref{def_Klnnp},
the $\Min$--$\Max$ terms stem
from the usual Legendre expansion of the Newtonian interaction potential.
The computation of ${ K_{n\np}^{\ell} }$ is the overall bottleneck
of the whole calculation of the \SRR\ diffusion coefficients which we have to address.
A naive inspection of equation~\eqref{def_Klnnp}
would lead us to believe that its computational complexity
scales like ${ \mO (K^{2}) }$,
with $K$ the number of sampling points
used to discrete both anomalies.
Fortunately, one can take inspiration from multipole
methods~\citep[see, e.g.\@,][]{Fouvry+2020}
to compute them much more efficiently,
yielding a computational complexity
scaling like ${ \mO (K) }$.
This is detailed in Appendix~\ref{sec:Multipole}.

Once the coupling coefficients have been estimated,
we rely on equation~\eqref{Djj_SRR} to evaluate
the diffusion coefficients.
This requires in particular to solve for the resonance condition
from equation~\eqref{resonance_constraint}.
For a given wire ${ (a,j) }$
and a given resonance pair ${ (n , \np) }$,
this amounts to finding all the wires ${ (\ap , \jp) }$
for which the resonance condition
${ n \nup (a,j) \!=\! \np \nup (\ap , \jp) }$ is satisfied.
We detail in Appendix~\ref{sec:ResCond}
our approach to solve the resonance condition,
improving upon the method from~\cite{BarOrFouvry2018}.
The performance of the code is given in Table~\ref{table:perf}.
The corresponding code is publicly available (see the data distribution policy below).

\begin{figure}
    \centering
   \includegraphics[width=0.45 \textwidth]{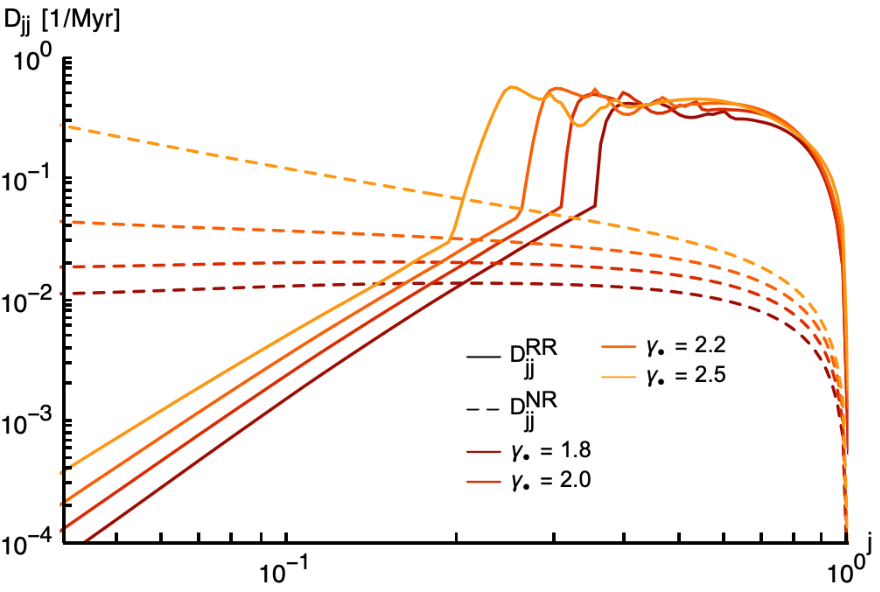}
   \caption{Illustration of the \RR\ (full line) and \NR\ (dashed line)
   diffusion coefficients. For the a Top-Heavy model
     (see section~\ref{sec:Model}), we vary $\gamma_{\bullet}$
     for a given value semi-major axis ${ a \!=\! 10 \, \mpc }$
     and the harmonic cutoff ${ \ellmax \!=\! 10 }$.
  The diffusion coefficients go to 0 as ${ j \!\rightarrow\! 1 }$
  (circular orbits), while the \RR\ ones get drastically reduced
  for very eccentric orbits~\citep{Merritt2011,BarOrAlexander2016}.
  As such, for small enough $j$, the \NR\ coefficients
  dominate over the \RR\ ones.
}
   \label{fig:DfixSMA}
\end{figure}

Figure~\ref{fig:DfixSMA}
 gives an example of a computation
of the \RR\ diffusion coefficients for a fixed
value of the semi-major axis.
In particular, we recover the drastic damping
of the \RR\ diffusion coefficients for very eccentric orbits.
This is due to the divergence of the relativistic precession
frequencies for ever more eccentric wires,
which prevents these wires from resonating
with the bulk of the other wires~\citep{Merritt2011,BarOrAlexander2016}.
As can be noted from Fig.~\ref{fig:DfixSMA},
very eccentric wires (${ j \!\sim\! 0 }$) are then immune to the \RR\ diffusion,
and can only keep diffusing under the effect of the \NR\ contributions.

\subsection{Non-resonant relaxation}
\label{sec:NR}

A second process through which test stars
relax in eccentricities originates from \NR\@~\citep[see \S{7.4.4} in][]{BinneyTremaine2008}.
In that case, it is the slow build-up of nearby scatterings
that ultimately drives the diffusion of their orbital parameters.

In order to evaluate the associated diffusion coefficient, ${ \DNR_{jj} (a,j) }$,
we used the exact same approach
as in Appendix~{C} of~\cite{BarOrAlexander2016}.
In a nutshell, the calculation proceeds as follows.
(i) At a given phase-space location ${ (\br , \bv) }$, one computes
the local velocity diffusion coefficients,
${ \langle \delta \bv \rangle (\br , \bv) }$
and ${ \langle (\delta \bv)^{2} \rangle (\br , \bv) }$,
see equation~{(7.83a)} in~\cite{BinneyTremaine2008}.
We note that here this calculation is greatly simplified
by our assumption that the background cluster
is fully relaxed, i.e.\ the cluster follows an isotropic \DF\@, ${ \Ftot \!=\! \Ftot (E) }$.
(ii) The local diffusion coefficients are then translated
into local diffusion coefficients in energy and angular momentum,
e.g.\@, ${ \langle \delta E \rangle (\br , \bv) }$, ${ \langle \delta J \rangle (\br , \bv) }$.
(iii) These local kicks then accumulate as the star moves along its Keplerian
wire. Following an orbit-average,
one obtains therefore the associated orbit-averaged diffusion coefficients,
e.g.\@, ${ \langle \Delta J \rangle (a ,j) \!=\! \!\oint \tfrac{\rd M}{2\pi} \langle \delta J \rangle }$.
(iv) Having obtained the first- and second-order diffusion coefficients
within the orbital coordinates ${ (E,J) }$,
we can obtain the associated diffusion coefficients
in the ${ (a,j)}$ space through the appropriate change of variables,
in particular
${ \langle (\Delta j)^{2} \rangle (a , j) \!=\! \DNR_{jj} (a,j) }$.
In practice, we define the Coulomb logarithm of a family
as ${\ln \Lambda_{i} \!=\! \ln (M_{\bullet}/m_{i})}$~\citep[see equation~{(7.84)} of][]{BinneyTremaine2008}.
Because they do not involve any resonance condition,
these \NR\ diffusion coefficients are numerically
much less demanding to compute than the \RR\ ones.

\begin{figure}
    \centering
   \includegraphics[width=0.5 \textwidth]{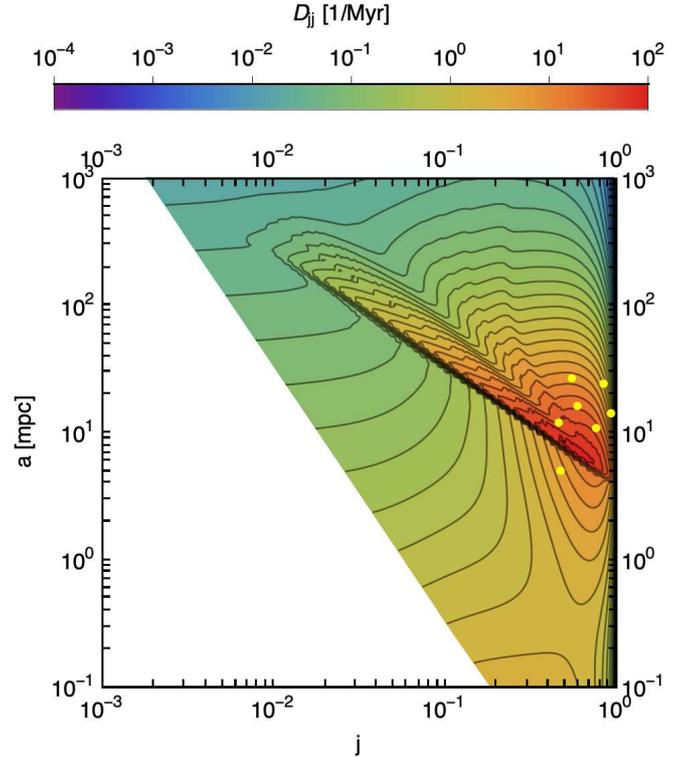}
   \caption{Illustration of the variation of the total diffusion coefficients, ${ D_{jj} (a,j) }$, in orbital space (semi major axis, $a$ and eccentricity, $j$) for the Top-Heavy model (see section~\ref{sec:Model}).
   Here are also represented in yellow all
   the S-stars~\citep{Gillessen2017},
   which were used to constrain the properties
   of the underlying unresolved stellar cluster.
   The
   white region on the left represent the location in orbital space
   of the central \BH\@'s loss cone.
   In the centre of
   the figure, resonant couplings in \RR\
   create this rugged but accurate aspect,
   that can be linked to that of the isocontours
   of the resonance frequencies, see Fig.~\ref{fig:Freq}.
   }
   \label{fig:DTriangle}
 \end{figure}

In Fig.~\ref{fig:DfixSMA}, we also illustrate
these \NR\ diffusion coefficients.
In practice, contrary to the \RR\ ones,
the \NR\ diffusion coefficients are mostly independent
of the stars' eccentricities.
Finally, in Fig.~\ref{fig:DTriangle}
we illustrate the overall dependence of the total diffusion coefficients
from equation~\eqref{sum_Djj},
i.e.\ both the \RR\ and \NR\ contributions,
in the whole ${ (a,j) }$ orbital space.
In that figure, one can clearly note
the presence of resonance lines associated with \RR\@.
One also notes that the bulk of the currently observed S-stars
lie in a region of orbital space,
where the diffusion of eccentricities is dominated
by resonant effects.
As a consequence, it is essential to account
for these resonant mechanisms
in order to accurately describe the dynamical fate
of the S stars' eccentricities.
One notes finally that the diffusion coefficient
varies significantly as a function of $j$
and stalls dramatically for ${ j \!\to\! 1 }$.
As a result, it takes a much shorter amount of time
for initially low eccentricity orbits to thermalise,
or equivalently for a given age,
it requires less massive unresolved perturbers
(see also Fig.~\ref{fig:LR_test_mass} below).

\section{Application}
\label{sec:Application}

Having quantified the two main diffusion processes
through which stars can relax in eccentricities
in galactic nuclei,
let us use them as dynamical probes
in the context of the recent observation
of the S-stars' eccentricities within SgrA*.

\subsection{Model's assumption}
\label{sec:Model}

For observational data,
we use the orbital parameters
listed in~\cite{Gillessen2017}.
Specifically, we use the ${(a,j)}$ coordinates
for seven of those stars
(S1, S2, S4, S6, S8, S9, S12).
Indeed, for these stars, \cite{Habibi2017} provides
us also with their main-sequence ages.
These ages are a measure of the total time
that the diffusion equation~\eqref{eq:FP_nice}
has had to operate.
For simplicity, we assume that on these timescales,
the \NR\ of the S-stars' energies
did not drive any significant diffusion,
so that the stars' semi-major axes, $a$, are kept fixed.
Regarding the initial conditions for the stars' eccentricities,
we investigate two possibles scenarii,
either originating from binary tidal disruptions~\citep{Hills1988, Gould_2003, Alexander2017},
i.e.\ large initial eccentricities,
or from an episode of disc formation~\citep{ALEXANDER200565, Levin2007,Koposov2020},
i.e.\ small initial eccentricities.
In practice, we assume that the S-stars
are initialised following a Gaussian distribution centered at ${ j (t \!=\! 0) \!=\! 0.2 }$ -- with width 
$0.02$ -- 
to mimick the eccentricity distribution of binary disruptions \citep{Generozov2020},
or ${ j (t \!=\! 0) \!=\! 0.9 }$
to mimic in-situ disc formation.
For alternative scenarii, see also \cite{Madigan2009, Perets2007}.

Let us also now make key assumptions
regarding the background old stellar cluster.
As detailed in Appendix~\ref{sec:Cusp},
we assume that it is composed of various sub-populations
of different individual masses $m_{i}$
with a total mass ${ M_{i}(<\! a_{0}) }$ enclosed within
a physical radius $a_{0}$.
In addition, we also assume that each population
follows a thermal distribution in eccentricity,
and infinite power-law distribution in semi-major axes,
that is,
\begin{equation}
M_{i}(<a) = M_{i}(<a_{0}) \bigg(\frac{a}{a_{0}}\bigg)^{3-\gamma_{i}} .
\label{Mass_distribution}
\end{equation}
We also assume here that throughout their eccentricity diffusion,
the S-stars are treated as test stars.
As such, they do not contribute to the system's mean potential,
and do not interact with one another. 

Since we expect the background to be thermal, the \RR\ dynamical friction vanishes exactly~\citep{BarOrFouvry2018}.
Conversely, we also neglect the \NR\ part of dynamical friction,
since energy diffusion is inefficient in quasi-Keplerian systems on \SRR\ timescale~\citep{BarOrAlexander2016}.

Assuming a two-family background
composed of stars and another heavy sub-population
(e.g.\@, \IMBHs\@),
we then have a total of 7 free parameters for the available models,
namely the power indices ${ (\gamma_{\star} , \gamma_{\bullet}) }$,
the individual masses ${ (m_{\star}, m_{\bullet} ) }$,
the total enclosed masses ${ (M_{\star} (<\!a_{0}), M_{\bullet} (<\!a_{0}) )}$
as well as the initial eccentricity of the S-stars, ${ j_{0} \!=\! j (t \!=\! 0) }$.
These models are complemented with the observed constraints
on the seven considered S-stars,
namely their main-sequence age,
as well as their observed semi-major axis and eccentricity.

In practice, we started our investigation
from the two-family Top-Heavy model of~\cite{Generozov2020}.
Using ${ a_{0} \!=\! 0.1} $ pc,
the fiducial model contains both stars and \IMBHs\ 
\begin{equation}
\begin{cases}
\displaystyle m_{\star} \!\!\!\!\! & = 1 M_{\odot} ,
\\
\displaystyle m_{\bullet} \!\!\!\!\! & = 50 M_{\odot} ,
\end{cases}
\begin{cases}
\displaystyle M_{\star} ( \!<\! a_{0}) \!\!\!\!\! & = 7.9 \!\times\! 10^{3} M_{\odot} ,
\\
\displaystyle M_{\bullet} (\!<\! a_{0}) \!\!\!\!\! & = 38 \!\times\! 10^{3} M_{\odot} ,
\end{cases}
\begin{cases}
\displaystyle \gamma_{\star} \!\!\!\!\! & = 1.5 ,
\\
\displaystyle \gamma_{\bullet} \!\!\!\!\! & = 1.8 ,
\end{cases}
\label{eq:TopHeavyModel}
\end{equation}
where the star parameters follow~\cite{Schodel2017}.
We note that such a model is compatible
with the current constraints associated
with S2's pericentre shift~\citep{Gravity2020},
since ${M_{\bullet} (\!<\! r_{\mathrm{apo}}^{\mathrm{S2}}) \!+\! M_{\star}(\!<\! r_{\mathrm{apo}}^{\mathrm{S2}}) \!\simeq\! 2500 M_{\odot}}$.

\subsection{Methodology}
\label{sec:Methodology}

Having picked a set of initial conditions for the S-stars,
and a model for the background clusters,
we are now in a position to compute the associated diffusion coefficients.
In order to determine whether or not such a model
is compatible with the observational constraint
of a significant eccentricity relaxation of the S-stars,
we proceeded as follows.

We first compute the \RR\ and \NR\ diffusion coefficients
for the $a$ of the S-stars considered.
The total diffusion coefficients are then interpolated
and we integrate equation~\eqref{eq:FP_nice}
forward in time using finite elements.
More precisely, we rely on the 
so-called {method-of-lines} implemented in the
\texttt{Mathematica}
\href{https://reference.wolfram.com/language/tutorial/NDSolveMethodOfLines.html}{\texttt{NDSolve}} function, which
discretizes the $j$ dimension and integrates the
semi-discrete problem as a system of Cauchy's ODEs.
As the semi-major axes are conserved,
they can be integrated separately.
As such, we integrate
equation~\eqref{eq:FP_nice}
for each of the seven considered S-stars,
for a total time equal to the age of the star. In order to ensure that the \PDF\ stays
normalised during the integration, it is useful to rewrite equation~\eqref{eq:FP_nice} into
\begin{equation}
j^{2}\frac{\partial P}{\partial t} = \frac{j^{2}}{2} \frac{\partial }{\partial j} \bigg( D_{jj} \frac{\partial P }{\partial j} \bigg)-\frac{1}{2} \bigg[ j \frac{\partial (D_{jj} P)}{\partial j} \bigg] +\frac{D_{jj} P}{2} ,
\label{eq:FP_nice_renorm}
\end{equation}
in order to avoid the $1/j$ singularities.
In practice, we also checked the sanity of this integration
using stochastic Monte-Carlo realisations,
see Appendix~\ref{sec:MonteCarlo}.
Once these integrations performed,
we compare the reached \PDF\ to the
observed data of the S-cluster,
determining whether or not the background model
allowed for an efficient enough relaxation
of the S-stars eccentricities.

Let us denote a model with $\balpha$,
i.e.\ the collection of the seven parameters of the background clusters
and the S-stars' initial eccentricities.
We then define a model's likelihood as
\begin{equation}
L(\boldsymbol{\alpha}) = \prod_{k} P (j_{k} \,|\, a_{k}) ,
\label{Likelihood}
\end{equation}
where ${ k \!=\! 1,...,7 }$ go through the 7 S-stars mentioned before.
Relying on equation~\eqref{Likelihood},
we can then explore the space of parameters $\balpha$ and compare the various models to one another.
To that end, we use the \LR\ test through
\begin{equation}
\lambda_{R}(\balpha) = 2 \ln \bigg( \frac{L_{\rm{\max}}}{L(\balpha)} \bigg) \in [0,+\infty[ ,
\label{LR_test}
\end{equation}
When $\balpha$ maximises the likelihood,
 it minimises by definition this likelihood ratio (as would a $\chi^2$ analysis for Gaussian statistics),
such that ${ \lambda_{R}(\balpha) \!=\! 0 }$.
Then, we can reject a model $\balpha$
with confidence ${ 0 \!\leq\! p \!\leq\! 1 }$,
if the corresponding \LR\@, ${ \lambda_{R}(\balpha) }$,
lies above a certain (explicit) value $\eta_{p}$.
This is further detailed in Appendix~\ref{sec:LR_test}.

\subsection{Results with existing data}
\label{sec:Results}

\begin{figure*}
    \centering
          \includegraphics[width=0.45 \textwidth]{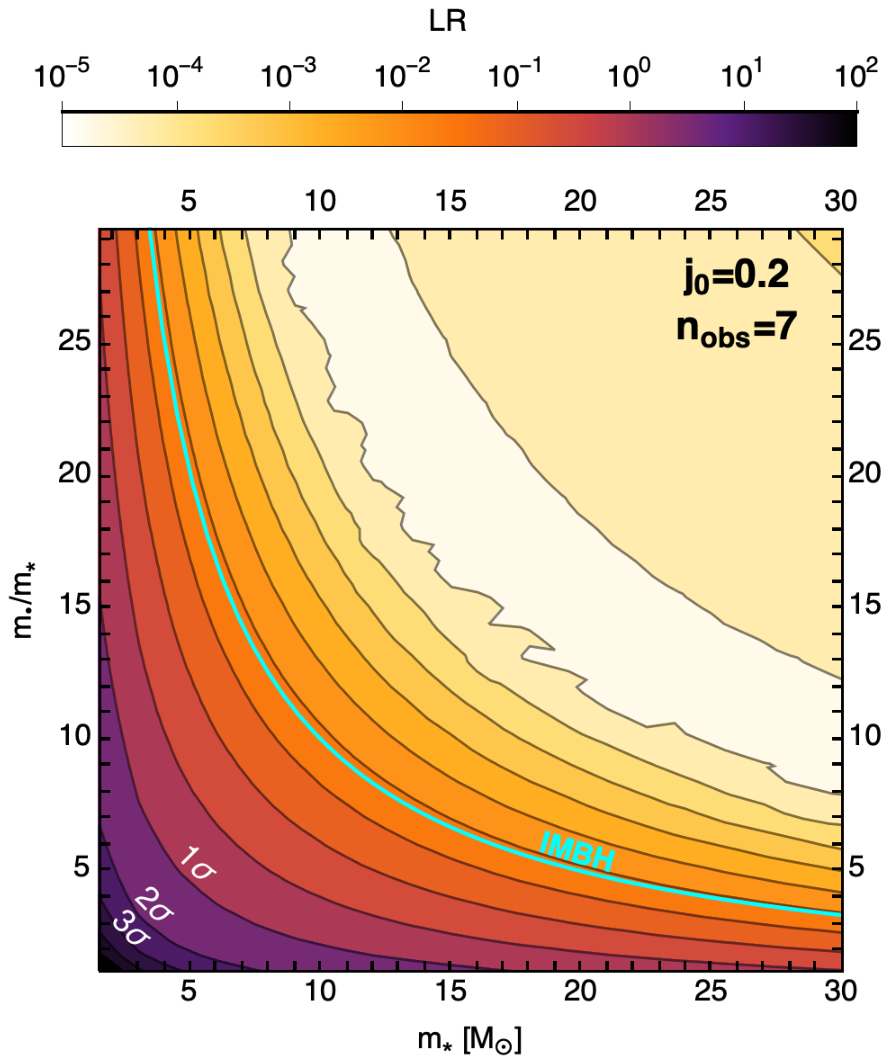}
                \includegraphics[width=0.45 \textwidth]{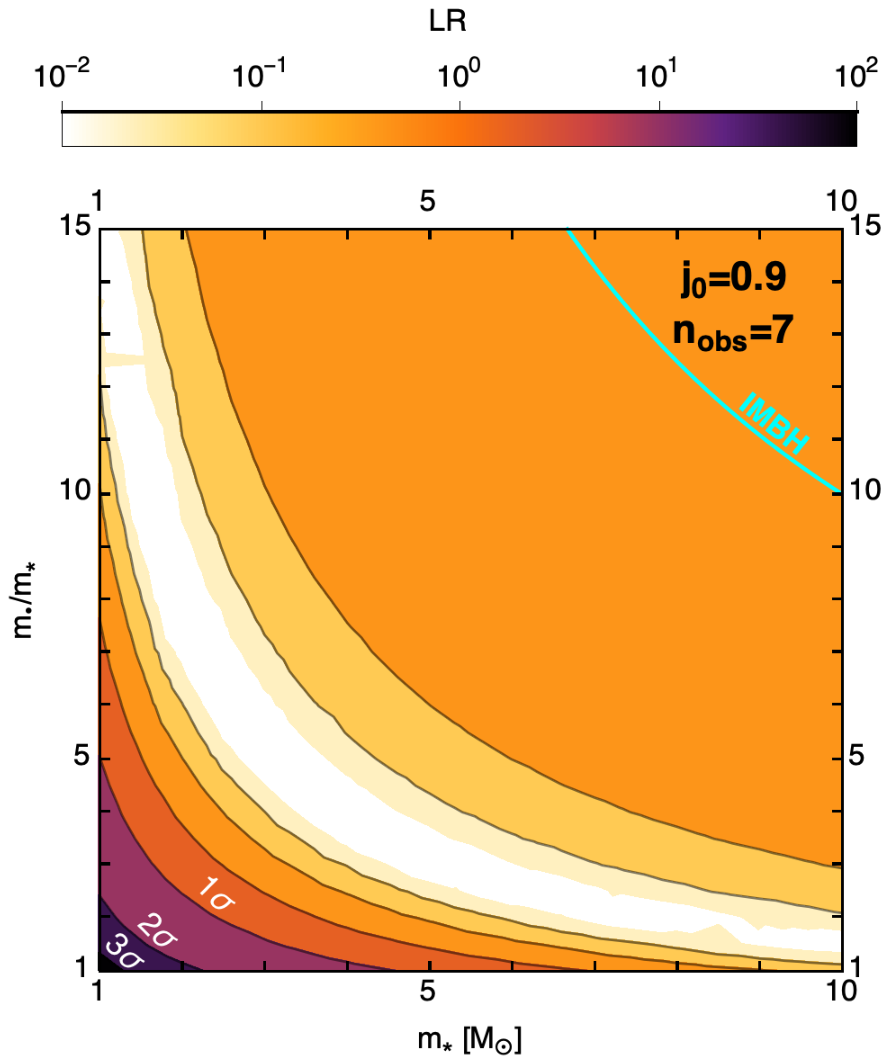}
            \includegraphics[width=0.45 \textwidth]{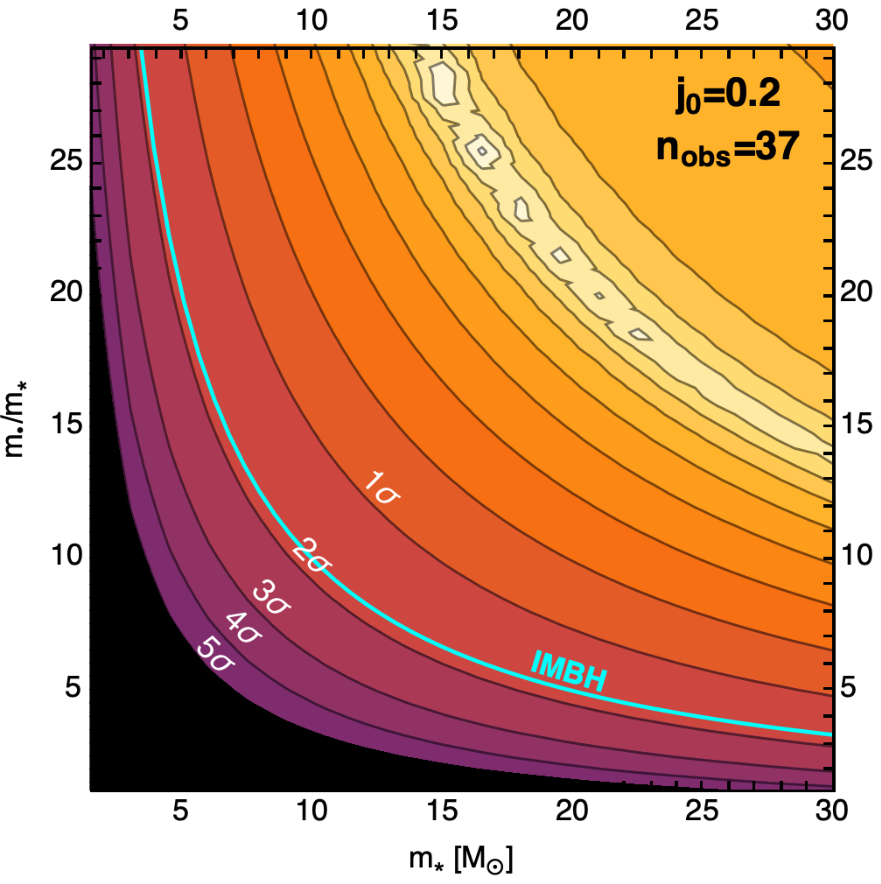}
            \includegraphics[width=0.45 \textwidth]{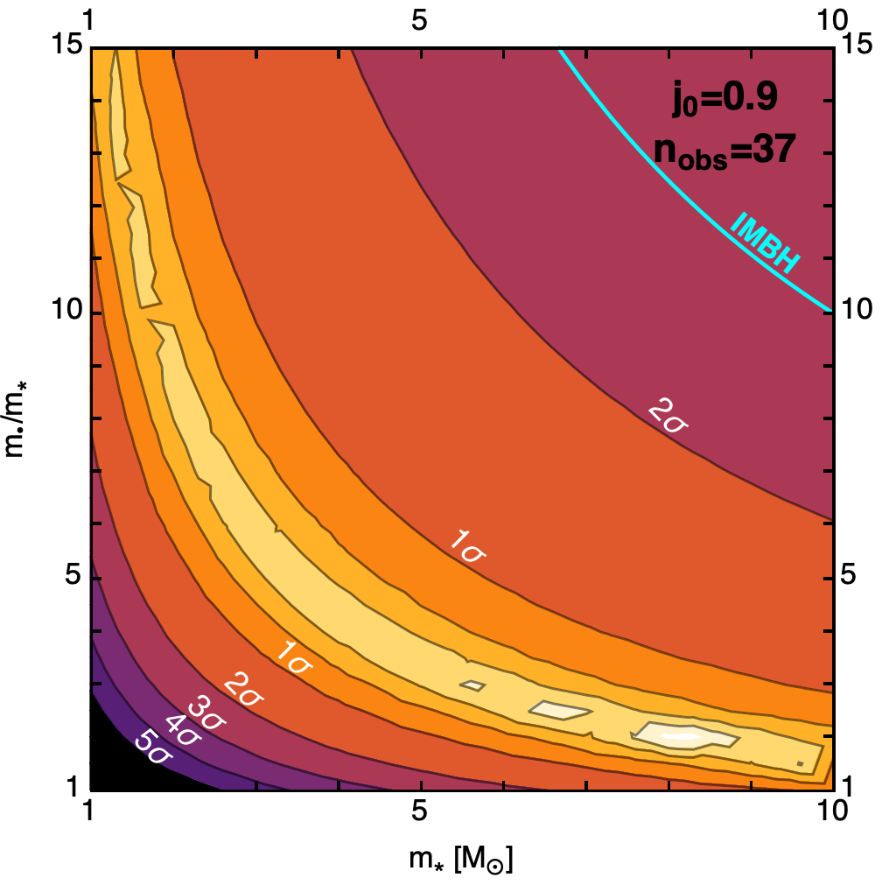}
   \caption{Confidence regions for
   the Top-Heavy model (see section~\ref{sec:Model}) using the maximum likelihood method applied to the observed S-stars,
   assuming a large initial eccentricity (${ j_0 \!=\! 0.2 }$, the canonical value, left panels),
   or a small initial eccentricity (${ j_0 \!=\! 0.9 }$, right panels).
   The cyan line corresponds to ${ m_{\bullet} \!=\! 100 \, M_{\odot} }$,
   above which the heavy objects are usually considered as IMBHs.
   The cusp's indices and the total enclosed masses
   are fixed to their fiducial values (see equation~\eqref{eq:TopHeavyModel}),
  but we let the individual masses, ($m_{\star}$, $m_{\bullet}$), vary.
  Confidence levels are inferred from the \LR\ test,
  see equation~\eqref{LR_test}.
  The top panels only used the 7 S-stars with known orbital
  parameters and stellar ages~\citep{Habibi2017},
  while the bottom ones expanded this observed sample
  using the other $30$ S-stars~\citep{Gillessen2017},
  assuming a common age ${ T \!=\! 7.1 \, \Myr }$,
  i.e.\ the average age of the constrained 7 S-stars.
  As expected, the smaller $j_{0}$,
  i.e.\ the more eccentric the stellar initial conditions,
  the slower the relaxation of the S-stars.
  Similarly, the larger the observed sample,
  the tighter the constraints on the background clusters.
  }
   \label{fig:LR_test_mass}
 \end{figure*}

As an illustration of the present method,
we first consider the Top-Heavy model
from equation~\eqref{eq:TopHeavyModel},
and let the individual masses $m_{\star}$ and $m_{\bullet}$ vary,
with the natural constraint ${ m_{\bullet} \!\geq\! m_{\star} }$
while fixing the total enclosed masses
${ M_{\bullet}(<\!a_0) }$ and ${ M_{\star}(<\!a_0) }$.
This is presented in Fig.~\ref{fig:LR_test_mass}.
In that figure, a model outside of the region
 of confidence ${n \sigma }$
 means that it can be discarded with confidence
 ${ n \sigma }$,
 as it would not allow the diffusion process to be fast enough to
 reach the observed eccentricity distribution
of the S-cluster.

As expected, in Fig.~\ref{fig:LR_test_mass} (top-left panel),
we recover that the larger the individual masses, 
the larger the underlying Poisson shot noise,
and therefore the more efficient the diffusion process,
and the faster the relaxation of the S-stars.
Conversely, Fig.~\ref{fig:LR_test_mass}
shows that models with small individual masses
cannot explain the current S-cluster's angular momentum \PDF\@.
As such, a relatively
massive set of background sources orbiting within the S-cluster is required to
 trigger a fast enough orbital diffusion of the observed stars over their lifetime.
Using the same data, in Fig.~\ref{fig:LR_test_mass} (top-right panel),
we also changed the initial eccentricity of the S-stars
to ${ j_{0} \!=\! 0.9 }$, to mimic an episode of disc formation.
As already observed in Fig.~\ref{fig:DTriangle},
we note that the diffusion coefficient is larger at smaller eccentricities,
so that the diffusion proceeds more swiftly,
hence enhancing the overall efficiency of the relaxation of the S-stars.

The global shape of the likelihood contours 
presented in Fig.~\ref{fig:LR_test_mass}
clearly illustrates the known dynamical degeneracy
in flipping \IMBHs\ and stars of the same mass,
as the  efficiency of eccentricity relaxation
is directly connected to the amplitude of the Poisson fluctuations
generated by the background clusters as a whole.
Interestingly, we note that all likelihood landscapes
presented in the top panels present an absolute minimum.
This suggests that, having only diffused a finite time,
the observed eccentricity distribution of the S-stars
is not fully thermal.

In order to increase the observed stellar sample,
and tighten the inferred model constraints,
we present in the bottom panels of Fig.~\ref{fig:LR_test_mass}
the same measurement but using $30$ additional
S-stars~\citep[as in Fig.~{13} of][]{Gillessen2017}.
Their individual ages was fixed to ${ T \!=\! 7.1 \, \Myr }$,
i.e. the average age of the 7 S-stars 
whose ages have been measured~\citep{Habibi2017}.
As expected, we recover that a larger sample of observed stars
leads to narrower contours around the likelihood extremum,
making the presence of second population of massive objects
all the more mandatory.
Finally, we also note that since the expanded sample of ${37}$ stars
contains stars with semi-major axes larger
than that of the initial 7 S-stars,
i.e.\ stars whose eccentricity relaxation is longer,
the location of the likelihood maximum
gets displaced to larger masses
as one increases the observed stellar sample.

\subsection{Prospective}
\label{sec:prospective}

Let us now carry out an experiment where we vary
the number of stars for which orbital parameters are available,
i.e.\ a prospective experiment appropriate for future surveys~\citep{Do2019}.

We consider a similar model as the one in Eq.~\eqref{eq:TopHeavyModel} where we set ${ m_{\star} \!=\! 5 \, M_{\odot} }$
and ${ m_{\bullet} \!=\! 20 \, M_{\odot} }$.
We now wish to probe how the number of observed stars impacts
our constraints on the determination of the background cluster parameters.
To that end, we take the same 7 S-stars as in Fig.~\ref{fig:LR_test_mass},
and consider their semi-major axes and main-sequence ages.
For each of the 7 semi-major axes,
we evolve the \PDF\ from Eq.~\eqref{eq:FP_nice}
from ${ j_{0} \!=\! 0.2 }$ for the entire star's observed lifetime.
From the resulting \PDFs\@, we draw $N$ stars
for each semi-major axis. In total, we therefore assume
that our observation sample is composed of a total of
${ n_{\mathrm{obs}} \!=\! 7N }$ stars.
This sample constitutes our mock data,
to which we apply the previous likelihood analysis.

\begin{figure}
    \centering
   \includegraphics[width=0.45 \textwidth]{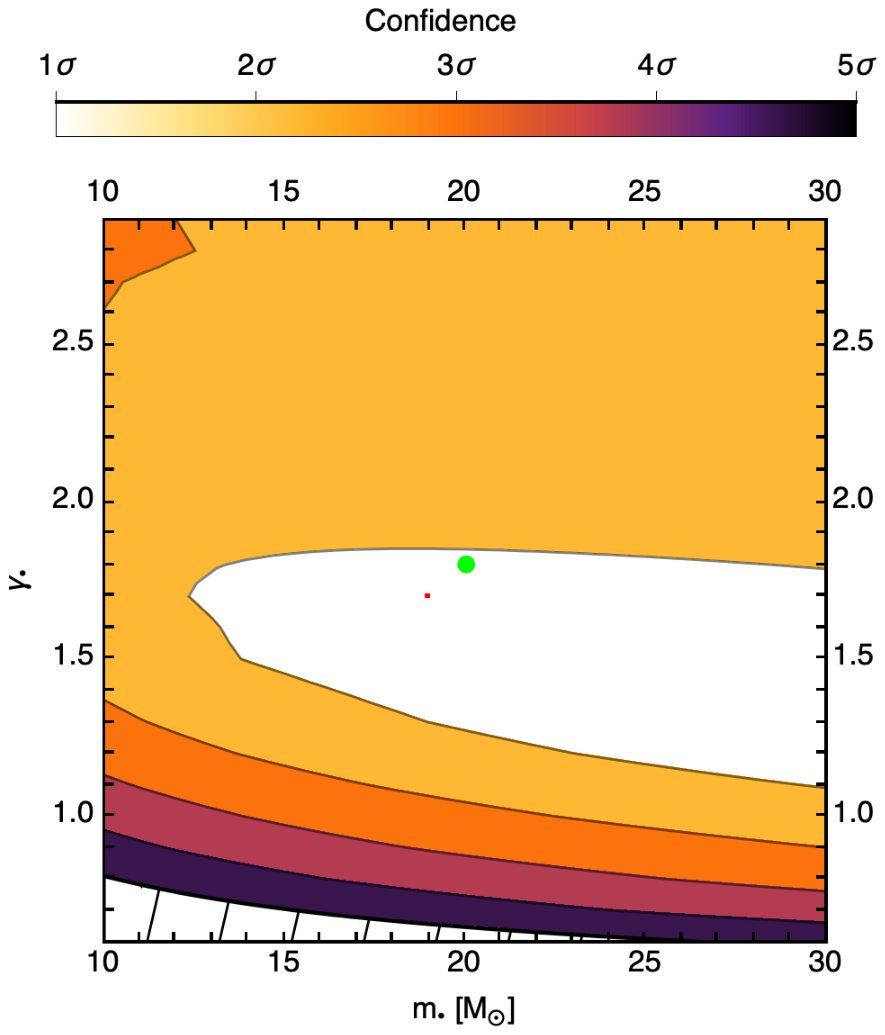}
      \includegraphics[width=0.45 \textwidth]{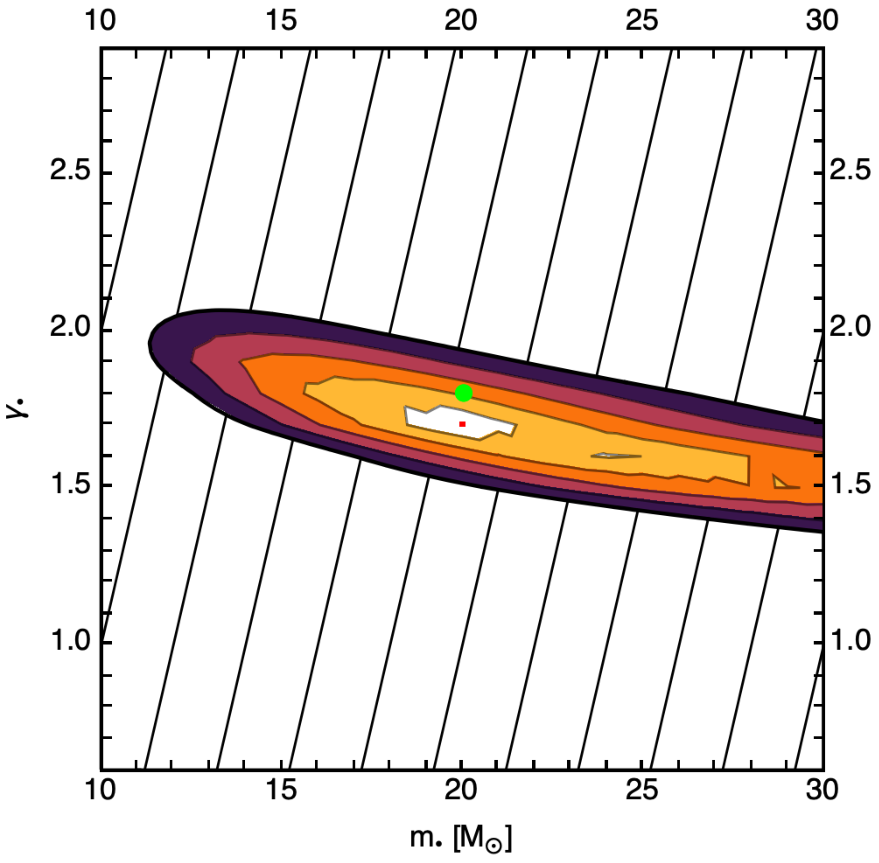}
   \caption{Same analysis as in Fig.~\ref{fig:LR_test_mass},
   but applied to mock data, as detailed in section~\ref{sec:prospective},
   as one varies the parameters, ${ (m_{\bullet}, \gamma_{\bullet}) }$,
   of the \IMBH\ population
   (keeping the total enclosed mass fixed).
   The top panel corresponds to mock data with ${ \nobs \!=\! 7 }$ stars,
   while the bottom panel uses ${ \nobs \!=\! 700 }$ stars.
   As expected, increasing the observed sample
   narrows the confidence contours around the maximum likelihood estimator (red dot),
   which converge towards the fiducial model (green dot). 
     }
   \label{fig:AdHoc_M_v_Gamma_0.2}
 \end{figure}

Following this approach, Fig.~\ref{fig:AdHoc_M_v_Gamma_0.2} shows
the ability of the  method to constrain
the parameters of the \IMBH\ population
given a larger mock sample.
While the stellar parameters ${ (m_{\star} , \gamma_{\star}) }$
are  observables,
we illustrate in that figure how the maximum likelihood
approach indeed allows us 
to constrain the parameters of the invisible dark cluster
${ (m_{\bullet} , \gamma_{\bullet}) }$,
that cannot be directly observed.
As ${ (m_{\bullet} , \gamma_{\bullet}) }$ are not degenerate
with one another, an increase in the number of
measured eccentricities (from ${ N \!=\! 1}$ to ${ N \!=\! 100 }$
from top to bottom panels)
narrows the confidence contours around the extremum of the likelihood,
which itself converges to a specific pair ${ (m_{\bullet}, \gamma_{\bullet} ) }$
close to the fiducial one (green dot).

\begin{figure}
    \centering
   \includegraphics[width=0.45 \textwidth]{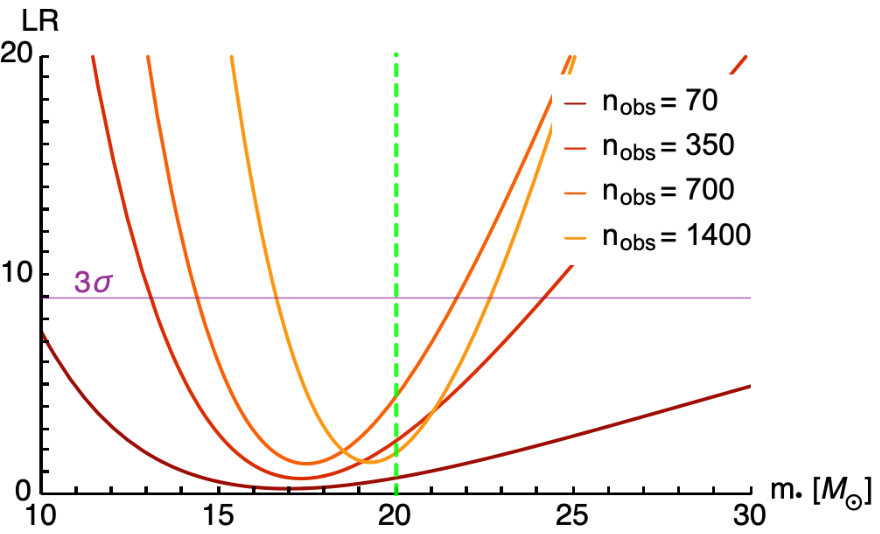}
      \includegraphics[width=0.45 \textwidth]{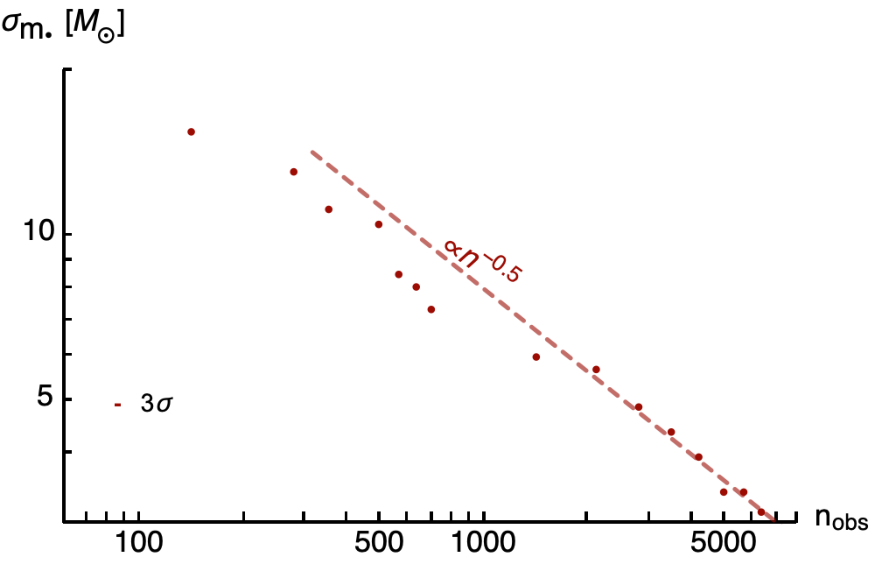}
   \caption{Same as in Fig.~\ref{fig:AdHoc_M_v_Gamma_0.2},
   using the same mock data,
   but with varying numbers of observed stars, $\nobs$.
   The top panel shows the evolution of the \LR\
   at fixed ${ \gamma_{\bullet} \!=\! 1.8 }$,
   as a function of $m_{\bullet}$,
   for various values of $\nobs$.
   The vertical green line represents the fiducial parameters model,
   ${ m_{\bullet} \!=\! 20 \, M_{\odot} }$.
   The horizontal line represents the value of the \LR\
   for a ${ 3 \sigma }$ confidence contour.
   The bottom panel illustrates the evolution of the accuracy, $\sigma_{m_{\bullet}}$,
   of the inferred \IMBH\ mass as a function of the $\nobs$,
   and for the $
   3\sigma$ confidence levels.
   For the modified TopHeavy model we use,
   we expect ${ \sim 10^{3} }$ stars within $5$ and ${ 20 \, \mpc }$.
   Of course we do not expect to observe so many stars around SgrA*,
   i.e.\  the asymptote $1/\sqrt{\nobs}$ will not be attained,
   because of crowding
   and, more importantly, because the total number
   of observed S-stars will be much smaller than 5000.
   }
     \label{fig:accuracy_v_n}
 \end{figure}

We further pursue this experiment in Fig.~\ref{fig:accuracy_v_n},
where we investigate the expected improvements
in the inferred constraints as a function
of the number of observed stars, $\nobs$.
For a given mock realisation,
we compute the uncertainty ${ \sigma_{m_{\bullet}} }$,
defined as the width of the \LR\ w.r.t. $m_{\bullet}$
at the ${3\sigma}$ height and fixed ${ \gamma_{\bullet} \!=\! 1.8 }$.
This is represented in Fig.~\ref{fig:accuracy_v_n}
as a function of $\nobs$.
Since the maximum likelihood estimator is asymptotically normal
and efficient~\citep[see, e.g.\@,][]{Wassermann04}
 it reaches the Cram\'er--Rao bound in the large $\nobs$ limit,
 so that 
 ${ \sigma_{m_{\bullet}} (\nobs) \! = \! \sigma_{3}/\sqrt{\nobs} }$,
with ${ \sigma_{3} \!\simeq\! 220 \, M_{\odot} }$.
Assuming crudely that the number of
resolved stellar orbits is proportional
to the survey's bolometric limit,
one can directly connect a target accuracy
with the survey's limiting magnitude.
Indeed, the survey's magnitude would simply
read ${ M \!=\! -2.5 \log_{10} [(\sigma_{m_{\bullet}} / \sigma_{3})^{-2}] }$.
Gaining a factor two in the accuracy of the mass
(i.e.\ ${ \sigma_{m_{\bullet}} \!\rightarrow\! \sigma_{m_{\bullet}} / 2 }$)
would require a survey that is at least ${ \Delta M \!=\! - 5 \log_{10} (2) \!\simeq\! -1.5 }$
magnitudes fainter.
Undoubtedly, upcoming surveys of SgrA*'s stellar neighbourhood, such as
GRAVITY+~\citep{Gravity+2019,Gravity2021},
TMT~\citep{Do2019}, 
and ELT/MICADO~\citep{davies2018micado,pott2018micado}
are on the verge of putting ever more stringent dynamical constraints
on the unresolved dark cluster.
Indeed, the central stellar cusp around SgrA*
is strongly confusion-limited for current observations on 8m class telescopes
with adaptative optics,
limiting in effect the reliable detection and measurement of positions of stars
to K magnitudes ${ \sim 16 \!-\! 17.5 }$, i.e.\ main-sequence B stars.
The combination of MICADO and the ELT will push the effective stellar detection sensitivity by ${ \gtrsim 5 }$ magnitudes with modest integration times~\citep{fiorentino2019maory}.

\section{Discussion and conclusion}
\label{sec:conclusion}

In the spirit of \cite{Generozov2020}, this paper was an attempt
at using kinetic theory and its dynamical diagnostics
to assess the structure of galactic nuclei.
We showed how 
 eccentricity diffusion in galactic nuclei can be used
to place constraints on the stellar and putative dark clusters present therein.
The recent observations
of the (quasi-) thermal distribution of eccentricities of the S-stars
orbiting SgrA*, in conjunction with updated computations
of the eccentricity diffusion coefficients,
can now be leveraged to this purpose.
Investigating a simple two-populations model
(see section~\ref{sec:Model}),
we showed how the presence of a heavy sub-population,
e.g.\@, \IMBHs\@, is mandatory to source an efficient
enough relaxation of the S-stars' eccentricities.
We jointly showed how only some ranges
of dark cusp's power law indices and masses are compatible
with that same dynamical constraint.
As expected, our analysis highlighted
intrinsic dynamical degeneracies
in permuting the visible and dark cluster.
Assuming that upcoming experiments
will better qualify the properties of the visible cluster,
kinetic theory will allow for dynamical dark matter experiment
to constrain both the typical mass and geometry of the \IMBH\ cluster. 

Finally, a simple fiducial experiment allowed us to quantify 
 the depth that upcoming surveys should achieve in order to 
e.g. double the accuracy on the \IMBH\@'s mass required to 
match the data.
More generally, this investigation suggests that it will be of interest to lift some of the degeneracies by increasing the number of measured stellar ages, better quantify the mass function and shape of the observed stellar cluster and initial eccentricity distribution, so that kinematic modelling
can further focus on dynamically
quantifying the properties of the dark cluster.

\subsection{Perspectives}

Let us now discuss some venues for future developments.
As shown in section~\ref{sec:Model},
the present investigation relies on various assumptions,
some of which one could hope to partially lift.
Our models for the old stellar and dark cluster
remain simplistic, and it will be worthwhile
to investigate possible contributions from other
populations such as a dark-matter dominated components,
or additional populations of \IMBHs\@.
Similarly, as already emphasised in equation~\eqref{eq:FP_nice},
we assumed that the background cluster
is spherically symmetric.
Yet,~\cite{SzolgyenKocsis2018} recently showed
that in systems with a large mass spectrum,
e.g.\ containing \IMBHs\@,
one could expect \VRR\ to lead to equilibria distribution
where the massive components follow a strongly anisotropic
structure, i.e.\ aligned within the same disc.
Such a structure could definitely affect the efficiency
of eccentricity relaxation within it.
Any additional non-trivial structures present in that \PDF\@,
e.g.\ non-spherically symmetric distributions
or dearth of stars in orbital space,
would also have to be explained by the present diffusion processes.
Similarly, on scales even closer to the central \BH\@,
we would also have to account for additional
relativistic corrections stemming from it,
e.g.\@, effects associated with its spin.

Observations show that 7-10\% of the stars may have originated from an infalling population.
These stars display significant rotation \citep{Do2020}
and likely populate a disc.
The most direct impact of that disc would be to induce mean-field torquing on the orbital planes, but it might also impact later on
the eccentricities within the cluster.
Recently, \citet{Szolgyen2021} have investigated this effect numerically and found that the timescale for the eccentricity decrease is much shorter than Chandrasekhar's dynamical friction timescale.
This supports previous findings by~\citet{MadiganLevin2012} that resonant dynamical friction, driven by orbit-averaged torques,
dominates over ordinary non-resonant dynamical friction,
driven by nearby encounters,
and leads to eccentricity decrease for a co-rotating disk. 
From an analytical perspective, a possible venue would be to revisit
the present kinetic theory, while relying on a St\"{a}ckel description of the cluster's density, so as to keep it integrable and account for its flattening. This would clearly be an order of magnitude more complicated than the path chosen in this paper, as it would increase the dimension of action space to be considered, as well as require one to use elliptic coordinates \citep[even the linear-response of flattened systems has scarcely been investigated in the literature,][]{Robijn1995}.

We emphasise that the mass in the S-star cluster
is only a small fraction of the total enclosed mass
within 1 arcsec of the central black hole.
As such, it is unlikely that the S cluster itself
strongly disturbs
the background stellar distribution.
We also assumed here that this background cluster was thermal (${ F(j) \!=\! 2 j }$) hence fully relaxed.
In that limit, it does not drive any \RR\ dynamical friction~\citep{BarOrFouvry2018}. 
Should we lift this assumption, a more accurate modelling would include the coupling between both components of the cluster 
as a  two-populations model. This would require integrating the coupled set of kinetic equations in time,
rather than relying on a frozen Fokker-Planck approximation
for the diffusion coefficient. 
While this might be a worthwhile endeavour for upcoming data sets,
it is clearly beyond the scope of this first investigation.

When modelling the S-stars' dynamics,
we assumed that the semi-major axes
of the stars were fixed throughout the diffusion,
owing to the orbit-average.
While accounting for the contributions from the \NR\ diffusion
coefficients in $a$, it could be interesting
to investigate whether any additional diffusion
in $a$-space would affect the present constraints.
As already noted, the initial conditions of the S-stars,
e.g.\@, very eccentric vs.\ quasi-circular,
strongly affect the efficiency of their eccentricity relaxation (see Fig.~\ref{fig:LR_test_mass}).
In particular, one can expect that the distribution
of the S-stars in semi-major axes
also carries some information on their initial formation mechanism.

Here, we focused our interest
on the innermost S-stars (${ a \!\simeq\! 5 \mathrm{\,mpc} }$),
which are known to have partially relaxed in eccentricity.
This allowed us to place constraints on cluster models
so that admissible clusters have to
source an eccentricity diffusion that is fast enough.
One could use a similar approach
to investigate the relaxation of S-stars further out.
These outer stars have only very partially relaxed in eccentricity,
so that any admissible cluster model
must source a diffusion that is slow enough
for these outer regions not to have fully relaxed. 
Leveraging both constraints,
one should be in a position to effectively bracket
cluster models,
given that their induced diffusion must be both
efficient enough in the inner regions,
and inefficient enough in the outer ones. 
A same double-sided investigation
could also be carried out in the context of the \VRR\
of the same S-stars, as it has been observed that
the innermost stars follow a spherically symmetric distribution,
while the outer ones tend to be aligned within a disc~\citep{Bartko2009,Yelda2014},
i.e.\ orientation neighbours
have not been separated~\citep{Giral+2020}.
Once again, simultaneously accounting for all these dynamical constraints
will allow for better characterisations of SgrA*'s dark and visible structures.

Finally, future observations will undoubtedly prove
useful in placing these investigations on firmer grounds.
First, the interferometer GRAVITY
is currently tracking in details the trajectory of S2~\citep{Gravity2020}.
Any deviations of its orbit from S2's expected mean-field trajectory,
i.e.\ the expected Keplerian dynamics and in-plane precession,
will bear imprints from the fluctuations of the gravitational
potential on the scale of S2's orbit,
that kinetic theory should be able to describe.
Similarly, a possible observation from GRAVITY
of stars on scales even smaller than S2
would also carry essential information on SgrA*'s stellar structure
on smaller scales, i.e.\ closer to the central \BH\@.
On larger scales, one expects that observations
from upcoming thirty-meter telescopes~\citep{Do2019}
will allow for a finer characterisation
of the S-stars current distribution, ${ P (a,j,t) }$,
a very valuable dynamical information as shown in section~\ref{sec:prospective}.
In particular, the dependence of $P$ w.r.t.\ $a$
is strongly dependent on the formation mechanism of these stars.
Regarding the dependence w.r.t.\ $j$,
one could in particular hope to measure the scale,
i.e.\ the $a$, at which the S-stars diffuse less and less efficiently towards a
 thermal distribution of eccentricities, 
hence strongly constraining the efficiency of the diffusion mechanisms.
We note that the present maximum likelihood formalism
can naturally be extended to account for the measurement uncertainties,
such as on stellar ages.

Eventually, this line of investigation should prove useful in constraining super massive black hole formation scenarios.

\subsection*{Data Distribution}
The data underlying this article 
is available through reasonable request to the author.
The code is distributed on github at the following URL: \href{https://github.com/KerwannTEP/JuDOKA}{https://github.com/KerwannTEP/JuDOKA}.

\section*{Acknowledgements}

This work is partially supported by grant Segal ANR-19-CE31-0017
of the French Agence Nationale de la Recherche,
and by the Idex Sorbonne Universit\'e.
We thank St\'ephane Rouberol for the smooth running of the
Horizon Cluster, where the simulations were performed.
CP thanks Renaud Foy, Fabien Malbet and Eric Thi\'ebaut for (very) early discussions about this project.

\appendix

\section{Frequencies and resonances }
\label{sec:ResCond}

In the vicinity of a supermassive \BH\@,
Keplerian wires undergo an in-plane precession
of their pericentres,
as described by equation~\eqref{def_nup}.
In that relation,
the relativistic precession is given by
\begin{equation}
\nuGR (a , j) = 3 \frac{\rg}{a} \frac{1}{j^{2}} \nuKep (a) ,
\label{def_nuGR}
\end{equation}
where we introduced the (fast) Keplerian frequency,
${ \nuKep (a) }$, in equation~\eqref{def_nuKep},
as well as the gravitational radius ${ \rg \!=\! G \MBH / c^{2} }$.
In practice, this precession is said to be prograde
as one always has ${ \nuGR (a , j) \!>\! 0 }$.
The gravitational radius allows us to introduce a maximal eccentricity
\begin{equation}
\jlc (a) = 4 \sqrt{\frac{\rg}{a}} ,
\label{def_jlc}
\end{equation}
so that wires with ${ j \!\leq\! \jlc (a) }$ are assumed to be within the loss-cone~\citep{Merritt2013},
and, as such, are unavoidably absorbed by the central \BH\@.

In order to easily compute ${ \nuM (a,j)}$,
the mass precession frequency imposed by the background stellar cluster,
we assume that the stellar cluster follows
an infinite power-law distribution of the form ${ M (<\! a) \!\propto\! a^{3-\gamma} }$,
where ${ M(<\! a) }$ stands for the total stellar mass
physically enclosed within the radius $a$.
In that limit, following Appendix~{A} of~\cite{KocsisTremaine2015},
the mass precession frequency reads
\begin{equation}
\nuM (a , j) = \onuM (a) \, \hM (j) ,
\label{def_nuM}
\end{equation}
where in that expression, the dimensional dependence w.r.t.\ $a$ is captured by
\begin{equation}
\onuM (a) = \nuKep (a) \, \frac{M (<a)}{\MBH} ,
\label{def_onuM}
\end{equation}
while the dimensionless eccentricity dependence is given by
\begin{equation}
\hM (j) = \frac{j^{4 - \gamma}}{1 - j^{2}} \bigg[ P_{1 - \gamma} (1/j) - \frac{1}{j} P_{2 - \gamma} (1/j) \bigg] ,
\label{def_hM}
\end{equation}
with $P_{\alpha}$ the Legendre function of order $\alpha$.
In practice, near the edge ${ j = 1 }$, we note that ${ \hM (j) }$ can be advantageously replaced with
its Taylor expansion
\begin{equation}
\hM (j) \simeq \tfrac{1}{2} (-3 + \gamma) - \tfrac{1}{8} (-12 + \gamma + 4 \gamma^{2} - \gamma^{3}) (1 - j) ,
\label{expansion_hM}
\end{equation}
to avoid singularities.

Importantly, we note that the function ${ \hM (j) }$
is always negative for ${ \gamma < 3 }$.
Indeed, following equation~{(A2)} of~\cite{KocsisTremaine2015},
we can rewrite equation~\eqref{def_hM}
with the alternative integral form
\begin{align}
\hM (j) {} & = \frac{j^{2(3-\gamma)}}{\pi \sqrt{1-j^{2}}} \!\! \int_{0}^{\pi} \!\! \rd \psi \, \frac{\cos(\psi)}{(1+e \cos (\psi))^{3-\gamma}}\,,
\nonumber
\\
{} & = \frac{j^{2 (3 - \gamma)}}{\pi \sqrt{1 - j^{2}}} \!\! \int_{0}^{\pi/2} \!\!\!\!\!\! \rd \psi \, \cos (\psi)
\nonumber
\\
{} & \quad \times \bigg\{ \frac{1}{(1 \!+\! e \cos (\psi))^{3 - \gamma}} \!-\! \frac{1}{(1 \!-\! e \cos (\psi))^{3 - \gamma}}\bigg\} ,
\label{hM_integral}
\end{align}
which is explicitely negative for any potential satisfying ${ 3 \!-\! \gamma \!>\! 0 }$.
As a consequence, the mass precession is generically retrograde,
i.e.\ one has ${ \nuM (a , j) \!\leq\! 0 }$.

\begin{figure}
    \centering
   \includegraphics[width=0.45 \textwidth]{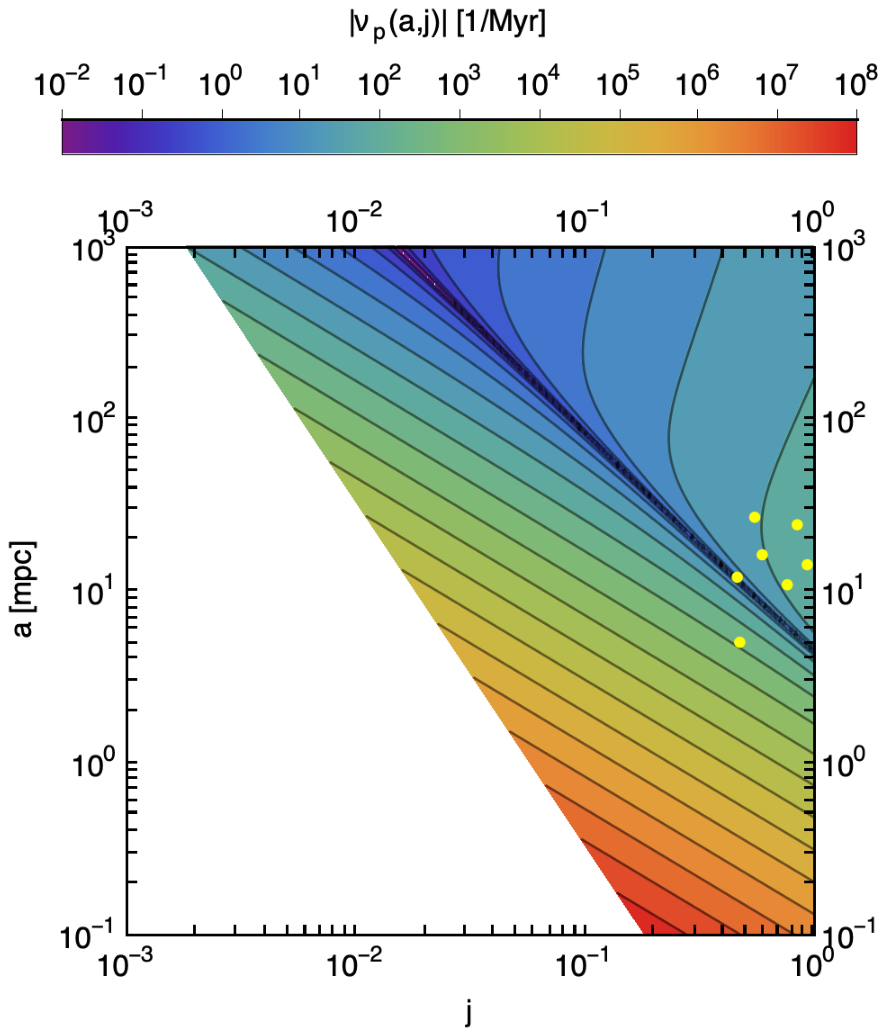}
   \caption{Illustration of precession frequencies,
     ${ |\nup (a,j)| }$, in orbital space for the model from Fig.~\ref{fig:DTriangle}.
     The orbital space locations of the seven S-stars used in our
     analysis are represented in yellow.
     For circular orbits and large semi-major axis,
     i.e.\ ${ j \!\to\! 1}$ and ${ a \!\gg\! 1 }$
   the precession is dominated by the mass precession
   and is therefore retrograde (${ \nup \!<\! 0 }$),
   while for low $j$ and low $a$,
   the precession is dominated by the relativistic precession
   and is therefore prograde (${ \nup \!>\! 0 }$).}
   \label{fig:Freq}
\end{figure}

Figure~\ref{fig:Freq}
illustrates the behaviour of the total precession
frequency, ${ \nup (a,j) }$,
as a function of the wire's underlying orbital parameters.
Lines of constant precession frequencies
correspond to the resonant lines along which
the \RR\ diffusion coefficients from equation~\eqref{Djj_SRR}
must be computed.
Note that the precession of very eccentric orbits
is dominated by the diverging relativistic corrections.
This is responsible for the ``Schwarzschild barrier''~\citep{Merritt2011,BarOrAlexander2016}
that explains the drastic reduction of the \RR\ diffusion
coefficients, shown in Fig.~\ref{fig:DfixSMA}.

\begin{figure}
    \centering
   \includegraphics[width=0.45 \textwidth]{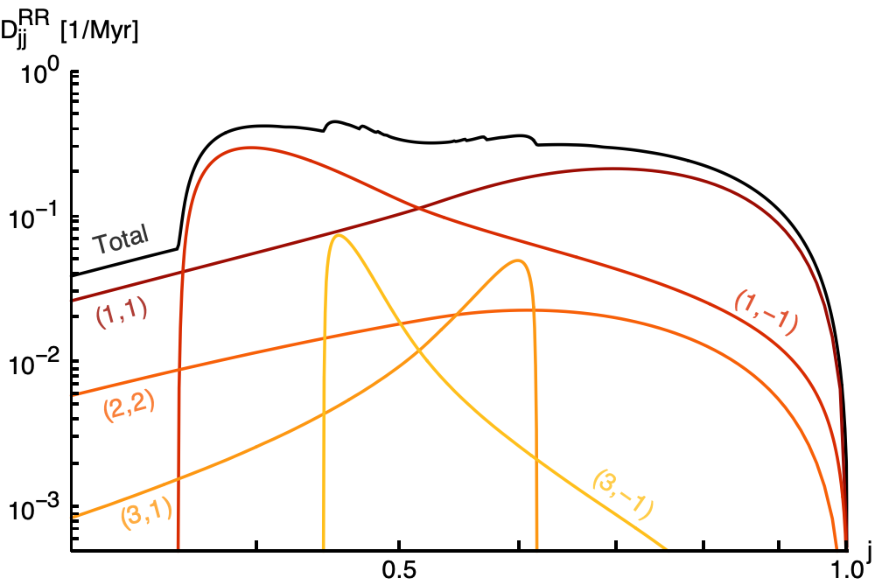}
   \caption{Illustration of the contribution to the \RR\
   diffusion coefficients of the different resonances ${ (n , \np) }$, 
     for a given value semi-major axis ${ a \!=\! 10 \mpc }$
     and the cutoff ${ \ellmax \!=\! 10 }$,
     for the model from Eq.~\eqref{eq:TopHeavyModel}.
     The \RR\ coefficients are typically dominated
   by ${ (n,\np) \!=\! (1,1) }$ (for small and large $j$)
   and ${ (n , \np) \!=\! (1 , -1) }$ (for intermediate $j$).
   For intermediate eccentricities, higher-order resonances
   also contribute.}
   \label{fig:DfixSMA_Resonances}
\end{figure}

In order to compute the resonant diffusion coefficients
from equation~\eqref{Djj_SRR},
we must solve the resonance condition
from equation~\eqref{resonance_constraint}.
For a given wire ${ (a,j) }$,
and a given resonance pair ${ (n , \np) }$,
this involves characterising all the wires ${ (\ap , \jp) }$
such that ${ \np \nup (\ap , \jp) \!=\! n \nup (a , j) }$,
i.e.\ identifying the appropriate level lines in Fig.~\ref{fig:Freq}.
In Fig.~\ref{fig:DfixSMA_Resonances},
we illustrate the contributions from the various resonance pairs
${ (n , \np) }$ to the total \RR\ diffusion coefficients.

Let us briefly detail our implementation for the search
of the resonant lines.
Here, the key remark is to note
that, following equations~\eqref{def_nuGR} and~\eqref{def_hM},
one always has ${ \rd \nup / \rd j \!<\! 0 }$.
As a consequence, for a given value of $\ap$,
it is straightforward to determine whether or not
there exists a $\jp$, with ${ \jlc (\ap) \!\leq\! \jp \!\leq\! 1 }$,
and ${ \np \nup(\ap , \jp) \!=\! n \nup (a , j) }$.
Using this approach, we may then identify
a domain ${ \apmin \!\leq\! \ap \!\leq\! \apmax }$,
within which the resonance condition can be satisfied,
by solving appropriately the resonance conditions
along the critical lines ${ j \!=\! \jlc (a) }$
as well as ${ j \!=\! 1 }$.
At this stage, we also enforce that ${ 16 \, \rg \!\leq\! \apmin }$
(see equation~\eqref{def_jlc})
as well as ${ \apmax \!\leq\! \rh }$,
with $\rh$ the considered influence radius
(e.g.\@, ${ \rh \!=\! 2 \mathrm{pc} }$ for SgrA*),
to ensure that we limit ourselves only to meaningful resonant regions
of orbital space.

Once the range ${ [\apmin , \apmax] }$ has been determined,
to emphasise the system's partial scale-invariance,
we sample this domain of semi-major axis
linearly in $\log$-space,
using ${ \Kres \!=\! 100 }$ points.
Finally, for a given value $\ap$ such that ${ \apmin \!\leq\! \ap \!\leq\! \apmax }$,
the associated resonant value $\jp$
is directly obtained by bisection.
For models with ${\gamma \!<\! 1.5}$,
it can happen that ${ a \mapsto \nup(a,j) }$
is not monotonic anymore for $j$ close to 1 (circular orbit),
leading to the possible appearance of a second range of semi-major axes
over which the resonance condition is satisfied.
When this is the case, we accordingly sample this domain
using the same method.

\section{Stellar cusps around SgrA*}
\label{sec:Cusp}

For the sake of simplicity,
we assume that all the background populations
follow infinite power-law distributions, which eases the resolution
of the resonance condition (see Appendix~\ref{sec:ResCond}).

Let us first specify our conventions for the normalisations
of their respective \DFs\@.
A given background population is characterised
by four numbers, namely $\gamma$,
the slope of the power-law profile,
$m$, the individual mass of the stars,
$a_{0}$, a given radius of reference,
and ${ M (<\! a_{0}) }$,
the total stellar mass physically within the radius $a_{0}$.
The number of stars per unit semi-major axis $a$
is then given by
\begin{equation}
N (a) = (3 - \gamma) \, \frac{N_{0}}{a_{0}} \, \bigg( \frac{a}{a_{0}} \bigg)^{2 - \gamma}\,.
\label{def_Na}
\end{equation}
In that expression, we introduced
${ N_{0} \!=\! g (\gamma) N (<\! a_{0}) }$ with
\begin{equation}
g (\gamma) = 2^{- \gamma} \, \sqrt{\pi} \, \frac{\Gamma (1 + \gamma)}{\Gamma(\gamma - \tfrac{1}{2})} ,
\label{def_g_gamma}
\end{equation}
where ${ N(<\! a_{0}) \!=\! M(<\! a_{0})/m }$
is the number of stars physically within a radius $a_{0}$.
This number should not be confused with $N_{0}$
that is the number of stars with a semi-major axis smaller than $a_{0}$.

In addition, we also assume
that each background population is thermally relaxed,
so that, as in equation~\eqref{def_Ftot},
we have
\begin{equation}
f (j \,|\, a) = 2 j ,
\label{relax_fj}
\end{equation}
which is the equilibrium solution of equation~\eqref{eq:FP_nice}.
When one accounts for the fact no wires can survive
within the loss-cone, this thermal \PDF\ gets truncated,
and becomes
\begin{equation}
f (j \,|\, a) = \frac{2 j}{1 - \jlc^{2} (a)} ,
\label{relax_fj_cut}
\end{equation}
where the limit eccentricity, ${ \jlc (a) }$,
is defined in equation~\eqref{def_jlc}.

\section{Coupling coefficients}
\label{sec:Multipole}
Let us now detail how one can efficiently
compute the coupling coefficients ${ K_{n\np}^{\ell} (a , j , \ap , \jp) }$
from equation~\eqref{def_Klnnp}.
When written explicitly, they read
\begin{equation}
K_{n \np}^{\ell} = \!\! \int_{0}^{2 \pi} \!\! \frac{\rd M}{2 \pi} \frac{\rd \Mp}{2 \pi} \cos (n f) \, \cos (\np \fp) \, \frac{\Min[r , \rp]^{\ell}}{\Max [r , \rp]^{\ell + 1}} ,
\label{rewrite_Klnnp}
\end{equation}
where $M$ and $\Mp$ stand for the mean anomalies of both orbits,
and we shortened the notation ${ K_{n\np}^{\ell} \!=\! K_{n\np}^{\ell} (a,j,\ap,\jp) }$.

First, we note that the function ${ r \mapsto r (M) }$ is an even function,
so that we can reduce the range of both angular integrals to ${ [0 , \pi] }$.
Moreover, in order not to have to invert Kepler's equation of motion,
it is more convenient to perform these integrals w.r.t.\ the true anomalies
$f$ and $\fp$.
In particular, the radius $r$ is directly obtained
from $f$ through~\citep{MurrayDermott1999}
\begin{equation}
r = \frac{a (1 - e^{2})}{1 + e \cos (f)} ,
\label{f_to_r}
\end{equation}
with the associated Jacobian
\begin{equation}
\frac{\rd M}{\rd f} = \frac{r^{2}}{a^{2}} \frac{1}{\sqrt{1 - e^{2}}} .
\label{Jac_M_f}
\end{equation}
Following these modifications,
we can rewrite equation~\eqref{rewrite_Klnnp} as
\begin{align}
K_{n\np}^{\ell} = {} & \!\! \int_{0}^{\pi} \!\! \frac{\rd f}{\pi} \frac{\rd \fp}{\pi} \, \frac{\rd M}{\rd f} \frac{\rd \Mp}{\rd \fp}
\nonumber
\\
{} & \times \cos (n f) \, \cos (\np \fp) \, \frac{\Min [r , \rp]^{\ell}}{\Max [r , \rp]^{\ell + 1}} .
\label{shorter_Klnnp}
\end{align}
At this stage, a naive approach would be to discretise
each integral into $K$ discrete steps,
and replace them with Riemann sums,
accounting for a total complexity in ${ \mO (K^{2}) }$.
Fortunately, dealing appropriately with the ratio
of $\Min$ and $\Max$,
equation~\eqref{shorter_Klnnp}
can be computed in ${ \mO (K) }$ operations,
as these integrals are almost separable.

First, we sample uniformly the integration intervals
from equation~\eqref{shorter_Klnnp} using $K$ nodes.
Specifically, we sample the true anomaly with
\begin{equation}
f_{k} = \Delta f \, \big( k - \tfrac{1}{2} \big)\,,
\quad \text{with} \quad
1 \leq k \leq K ,
\label{sample_f}
\end{equation}
where we introduced the step distance ${ \Delta f \!=\! \pi / K }$.
Here, following the midpoint-rule,
each sampling location is offset by a factor $\tfrac{1}{2}$.
This ensures that the ${2 \pi}$-periodic integrand
is sampled uniformly,
which allows for fast convergence of the result~\citep{Trefethen2014}.
Following this discretisation,
equation~\eqref{shorter_Klnnp} becomes
\begin{equation}
K_{n\np}^{\ell} = \frac{1}{K^{2}} \sum_{i , j} g_{i} \, \gp_{j} \, \frac{\Min [r_{i} , \rp_{j}]^{\ell}}{\Max [r_{i} , \rp_{j}]^{\ell + 1}} ,
\label{disc_Klnnp}
\end{equation}
where we introduced the function ${ g (r) \!=\! \cos (n f) \, \rd M / \rd f }$,
as well as the shorthand notations
${ g_{i} \!=\! g (r_{i}) }$ and ${ \gp_{j} \!=\! g(\rp_{j}) }$.
One can now use the particular structure of equation~\eqref{disc_Klnnp}
to drastically accelerate its evaluation.
To do so, we order the set of radii ${ \{ r_{i} , \rp_{j} \} }$ by increasing order.
We note that this can be done in ${ \mO (K) }$ steps,
provided that the two sets ${ \{ r_{i} \} }$ and ${ \{ \rp_{j} \} }$
are already ordered,
so that it only remains to merge the two lists.

\begin{figure}
    \centering
     \includegraphics[width=0.45 \textwidth]{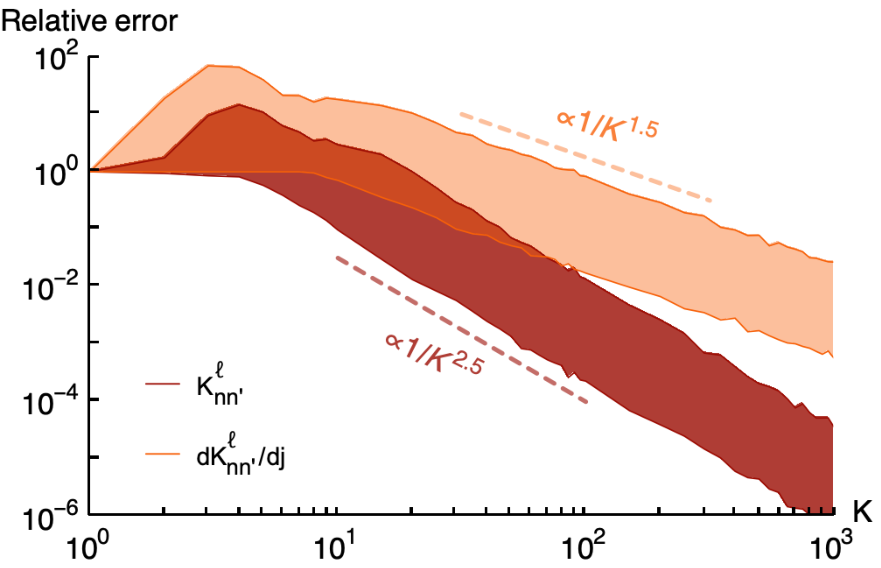}
   \caption{Illustration of the relative errors in the computations
     of ${ K_{n\np}^{\ell} }$ (in brown) and ${ \rd K_{n\np}^{\ell} / \rd j }$
     (in orange)
   using the multipole approach from equation~\eqref{disc_Klnnp_II},
   as a function of the number of nodes $K$. We approximated the real value by
   that obtained for ${ K \!=\! 2000 }$.
   The plotted regions represent the errors within the 16 and the 84 percentiles. The use of a mid-point rule allows
   for a fast convergence, though the regularity of the integrand does not
   allow for an exponential one.
   Errors are found to scale like ${ K \!\propto\! 1/K^{2.5} }$
   for $K_{n\np}^{\ell}$ and ${ K \!\propto\! 1/K^{1.5} }$
   for ${ \rd K_{n\np}^{\ell} / \rd j }$.
   In practice, we use ${ K \!=\! 100 }$ to compute $K_{n\np}^{\ell}$,
   with a relative error smaller than $1\%$.
   }
   \label{fig:ErrKlnnp}
\end{figure}

Following this sorting, we can now construct the array $w_{j}$
which, for ${ 1 \leq j \leq K }$, is defined as
\begin{equation}
w_{j} = \Card \bigg\{ i \in \{ 1,..,K \} \, \bigg| \, r_{i} \leq \rp_{j} \bigg\} ,
\label{def_w}
\end{equation}
with the boundary terms ${ w_{0} \!=\! 0 }$
and ${ w_{K+1} \!=\! K }$.
The double sum from equation~\eqref{disc_Klnnp}
can then be rewritten as
\begin{equation}
K_{n\np}^{\ell} = \frac{1}{K^{2}} \sum_{j = 1}^{K} \gp_{j} \, \big[ P_{j} + Q_{j} \big] .
\label{disc_Klnnp_II}
\end{equation}
In that expression, we introduced the reduced sums
$P_{j}$ and $Q_{j}$ that read
\begin{align}
P_{j} = \sum_{i = 1}^{w_{j}} g_{i} \, \frac{r_{i}^{\ell}}{r_{j}^{\prime \ell+1}} ;
\quad
Q_{j} = \sum_{\mathclap{i = w_{j} + 1}}^{K} g_{i} \, \frac{r_{j}^{\prime \ell}}{r_{i}^{\ell + 1}} .
\label{def_P_Q}
\end{align}
The key property here is note that the sum $P_{j}$
(resp.\ $Q_{j}$) can be computed in ${ \mO (K) }$
through an increasing (resp.\ decreasing) recurrence.
In order to highlight this property,
we define the partial sums
\begin{equation}
\delta P_{j} = \sum_{\mathclap{i = w_{j-1}+1}}^{w_{j}} g_{i} \, \frac{r_{i}^{\ell}}{r_{j}^{\prime \ell + 1}} ;
\quad
\delta Q_{j} = \sum_{\mathclap{i = w_{j}+1}}^{\mathclap{w_{j+1}}} g_{i} \, \frac{r_{j}^{\prime \ell}}{r_{i}^{\ell + 1}} .
\label{def_deltaP_deltaQ}
\end{equation}
The sums $P$ and $Q$ then satisfy the recurrence relations
\begin{align}
P_{1} = \delta P_{1} ; 
\quad
P_{j+1} {} & = \bigg[ \frac{\rp_{j}}{\rp_{j+1}} \bigg]^{\ell + 1} P_{j} + \delta P_{j+1} ,
\nonumber
\\
Q_{K} = \delta Q_{K} ;
\quad
Q_{j - 1} {} & = \bigg[ \frac{\rp_{j-1}}{\rp_{j}} \bigg]^{\ell} Q_{j} + \delta Q_{j-1} .
\label{rec_P_rec_Q}
\end{align}
Hence, given these two recurrence relations,
equation~\eqref{disc_Klnnp_II}
can be computed in ${ \mO (K) }$ operations.
Moreover, we note that the geometric prefactors
appearing in equation~\eqref{rec_P_rec_Q} are positive
and always smaller than $1$,
which helps ensuring the numerical stability of these recurrences.
In order to illustrate the quality of this discretisation scheme,
we present in Fig.~\ref{fig:ErrKlnnp}
the behaviour of the relative error in the computation
of $K_{n\np}^{\ell}$ as a function of $K$.
The relative errors appears to scale like ${ 1/K^{2.5} }$. From this observation, we can infer that ${ K \!=\! 100 }$
is enough to obtain a ${ 1\% }$ relative error for any orbital parameter.

Following equation~\eqref{fluctuation_dissipation},
we note that the computation of the first-order diffusion
coefficient ultimately also requires the computation
of ${ \rd K_{n\np}^{\ell} / \rd j }$.
It is straightforward to extend the previous recurrence relations
to compute such a derivative.
In Fig.~\ref{fig:ErrKlnnp},
we also illustrate the typical relative error
in the computation of ${ \rd K_{n\np}^{\ell} / \rd j }$.
In particular, we note that this gradient
introduces discontinuities in the integrand,
which reduces the convergence speed
of the method to an inverse power law proportional to $1/K^{1.5}$.

With such an approach, we expect that computing
the diffusion coefficient $\DRR_{jj}$ should have
a complexity linear w.r.t.\ $K$,
as recovered in Table~\ref{table:perf}.
\begin{table}
  \centering
\begin{tabular}{|c|c|c|c|c|c|c|c|c|}
   \hline
   $K$ &20 & 40 & 60 & 80 & 100 & 150 & 200 & 250 \\
   \hline
   Time (s) &0.51 & 0.72 & 0.96 & 1.13 & 1.30 & 1.89 & 2.44 & 2.96 \\
  \hline
  \hline
  $\ellmax$ & 6 & 8 & 10 & 12 & 14 & 16& 18 & 20 \\
   \hline
   Time (s) &0.38 & 0.76 & 1.35 & 2.12 & 2.93 & 4.16 & 5.54 & 7.72 \\
\end{tabular}
\caption{Computation time of ${ \DRR_{jj}(a,j) }$ for ${ a \!=\! 10 \mpc }$, ${ j \!=\! 0.6 }$ and for the same Top-Heavy model as in Fig.~\ref{fig:DTriangle}.
Fixing ${ \ellmax \!=\! 10 }$, we observe a linear complexity w.r.t.\ $K$ (second line),
while fixing ${ K \!=\! 100 }$, we observe a complexity in $(\ellmax)^{2.3}$
w.r.t.\ $\ellmax$.}
\label{table:perf}
\end{table}
Finally, the complexity of the computation of $\DRR_{j}$
w.r.t.\ $\ellmax$ follows a power law roughly proportionnal ${ (\ellmax)^{2.3} }$.
Such a scaling is primarily due to the growth
of the number of resonance pairs ${ (n , \np) }$
as $\ellmax$ increases.

\section{Simulating stochastic dynamics}
\label{sec:MonteCarlo}

One approach to simulate the relaxation
of the test stars' eccentricities
is to rely on Monte-Carlo realisations
of the underlying diffusion equation.
This is more easily done starting from
the traditional form of the \FP\ equation,
as given by equation~\eqref{eq:FP_classical},
which involves the first- and second-order diffusion
coefficients.
We used this alternative approach
to check the validity of our direct numerical integration
of the diffusion equation~\eqref{eq:FP_nice}.

\begin{figure}
    \centering
   \includegraphics[width=0.45 \textwidth]{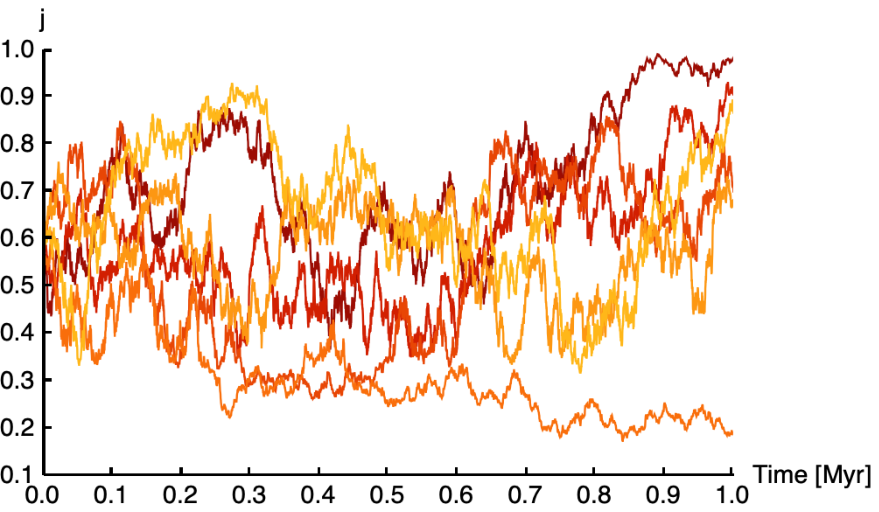}
   \caption{Illustration of stochastic random walks
     in eccentricities as driven by equation~\eqref{eq:Langevin_equation}
     with a timestep ${ \Delta t \!=\! 0.1 \, \kyr }$.
     Here, the test particles all have ${ a \!=\! 10 \mpc }$,
     are initialised with ${ j \!=\! 0.6 }$,
    and evolve within the same background model
     as in Fig.~\ref{fig:DTriangle}.
   }
   \label{fig:Stoch}
 \end{figure}

Following~\cite{Risken1989},
one can mimic the dynamics of a given test star through
the stochastic Langevin equation
\begin{equation}
\Delta j = D_{j} \, \Delta t + \sqrt{D_{jj} \, \Delta t} \, \xi (t) .
\label{eq:Langevin_equation}
\end{equation}
In that expression, ${ \Delta t }$ is our chosen (fixed) timestep,
and ${ \xi (t) }$ follows a normal distribution of unit variance,
uncorrelated in time.
Once one can simulate one realisation
of the stochastic dynamics,
we may use a large sample of test particles
to recover the time evolution of their smooth underlying
\PDF\@. 
In Fig.~\ref{fig:Stoch},
we illustrate some examples of random walks
in eccentricities.
To ensure that the random walks do not wander off
the range ${ j \in [0,1] }$,
we introduced a reflective barrier at ${ j \!=\! 0,1 }$.

Let us note that the stochastic walks
from Fig.~\ref{fig:Stoch}
do not describe any physically realistic random
walks on their own, but only in an average sense.
Indeed, here we have supposed that the ${\xi (t)}$ are
uncorrelated in time, whereas they are correlated
(at least on the fluctuations' coherence time)
in a real physical process.
However, their
average over realization accurately describes the evolution of the
corresponding \FP\ equation~\eqref{eq:FP_classical}.

\begin{figure}
    \centering
   \includegraphics[width=0.45 \textwidth]{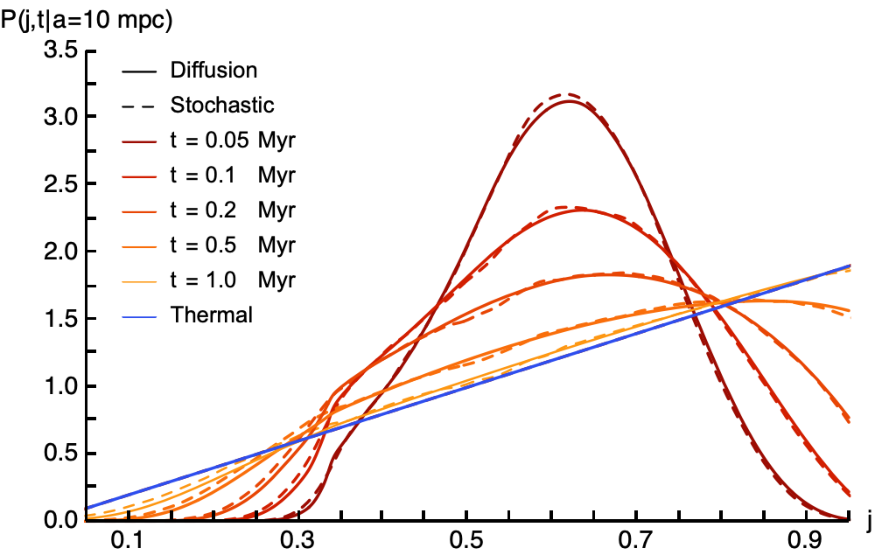}
   \caption{Comparison between
the direct integration of the diffusion equation~\eqref{eq:FP_nice} (full lines)
and its stochastic realisation through equation~ \eqref{eq:Langevin_equation} (dashed lines),
as a function of time,
and using the same initial conditions as in Fig.~\ref{fig:Stoch}.
For the stochastic evolution,
we considered a total of ${ N \!=\! 10^{6} }$ test particles,
evolved with the timestep ${ \Delta t \!=\! 0.1 \, \kyr }$.
Both approaches are found to be in very good agreement,
and ultimately relax, as expected, to the thermal distribution.
}
   \label{fig:NDSolve_v_Langevin}
 \end{figure}

Consequently, in Fig.~\ref{fig:NDSolve_v_Langevin},
we use ${ N \!=\! 10^{6} }$ test particles
to recover the \PDF\ at various times
and compare it with that obtained from the direct integration of the diffusion equation \eqref{eq:FP_nice} presented in the main text.
The two methods yield the same result which provides us with validation.
Furthermore, both methods also ultimately
asymptote to the full relaxation towards the thermal \PDF\@,
${ P_{\rm{th}}(j|a) \!=\! 2j }$.

\section{Likelihood and LR test}
\label{sec:LR_test}

Likelihoods measure the goodness
of fit of a statistical model to a data sample.
Its extremum, if it exists,
is associated with models
that extremise the probability of drawing
the observational sample at hand.
Given a a product of $K$ joint continuous \PDFs\@,
$P_{\balpha}(j \, | \, a_{k})$, depending on a parameter $\balpha$,
and a set of i.i.d. random sampling data $\{j_{i,k}\}$, we define the likelihood of a model as 
\begin{equation}
\mL (\balpha; \{j_{i,k}\}) = \prod_{k=1}^{K} \prod_{i=1}^{N} P_{\balpha}(j_{i,k}|a_{k}) .
\label{Likelihood_def}
\end{equation}
This allows us to define the likelihood ratio as
\begin{equation}
\lambda_{\balpha , N} = 2 \ln \bigg[ \frac{\mLmax ( \{j_{i,k}\})}{\mL (\balpha; \{j_{i,k}\})}\bigg] ,
\label{LR_def} \notag
\end{equation}
where $\mLmax$ corresponds to the maximum likelihood within the range of explored parameters.
The \LR\@, $\lambda_{\balpha , N}$,
is then a random variable,
that takes its values in ${ [0 , + \infty[ }$,
and depends on the model's parameters, ${ \balpha }$.

This \LR\ test allows us to compare models
with one another,
and discard those which are too unlikely.
Indeed, given a model $\balpha$, $\lambda_{\balpha , N}$
must fall close to $0$
for the corresponding model to drive efficiently
the eccentricity relaxation of the S-stars.
Given a confidence level ${ 0 \!\leq\! p \!\leq\! 1 }$,
we can define from it a confidence interval ${ [ 0,\eta_{p}] }$
within which $\lambda_{\balpha , N}$ must fall
for the model to be accepted.
Here, we choose ${ \eta_p }$ so that the probability $\mathbb{P}$ obeys
\begin{equation}
\mathbb{P}(0 \!\leq\! \lambda_{\balpha ,N} \!\leq\! \eta_{p}) = p .
\label{confidence_p}
\end{equation}
While we do not know the exact distribution of $\lambda_{\balpha , N}$,
owing to Wilks' theorem~\citep{Wilks1938},
$\lambda_{\balpha , N}$ converges to the
$\chi^{2}$ distribution as ${K \!\times\! N \!\rightarrow\! + \infty }$,
and $\eta_{p} $ obeys
\begin{equation}
\eta_{p} = 2 \big[\erf^{-1} (p)\big]^{2} .
\label{delta_p}
\end{equation}
In terms of the usual $\sigma$-levels of confidence,
since the Gaussian probability of being in the interval
${ [ - n \sigma, n \sigma] }$ is ${ p(n\sigma) \!=\! \erf(n/\sqrt{2}) }$,
then the corresponding threshold $\eta_{p (\! n \sigma \!)}$
simply becomes
${ \eta_{p (\! n \sigma \!)} \!=\! 2 [\erf^{-1} \erf (n/\sqrt{2})]^{2} \!=\! n^{2} }$.

In practice, to validate our calculations,
we also tried another non-parametric statistical estimator,
namely the Kolmogorov--Smirnov distance.
This led to the same conclusions.
In the main text,
we focused on the the maximum likelihood estimator,
because it converges asymptotically to a normal distribution
and is asymptotically efficient~\citep{Wassermann04},
leading to the ${ 1/\sqrt{\nobs} }$ behaviour
observed in Fig.~\ref{fig:accuracy_v_n}.

\vfill

\end{document}